# Thin-Film-Engineered Self-Assembly of 3D Coaxial Microfluidics with a Tunable Polyimide Membrane for Bioelectronic Power

Aleksandr I. Egunov[1]*, Hongmei Tang[1], Pablo E. Saenz[1], Dmitriy D. Karnaushenko[1], Yumin Luo[1,2], Chao Zhong[3], Xinyu Wang[3], Yang Huang[4], Pavel Fedorov[1], Leandro Merces[1,5], Minshen Zhu[1]*, Daniil Karnaushenko[1]* and Oliver G. Schmidt[1,6,7]*

[1] Research Center for Materials, Architectures and Integration of Nanomembranes, Chemnitz University of Technology, Chemnitz, Germany.

[2] Chair of Microsystems Technology, Ruhr University Bochum, Bochum, Germany

[3] Shenzhen Key Laboratory of Materials Synthetic Biology, Key Laboratory of Quantitative Synthetic Biology, Shenzhen Institute of Synthetic Biology, Shenzhen Institutes of Advanced Technology, Chinese Academy of Sciences, Shenzhen, China.

[4] Departament of Physics and Materials Science, City Univercity of Hong Kong, Hong Kong, China.

[5] Ilum School of Science, Brazilian Center for Research in Energy and Materials, Campinas, Brazil.

[6] Material Systems for Nanoelectronics, Chemnitz University of Technology, Chemnitz, Germany

[7] Nanophysics, Dresden University of Technology, Dresden, Germany

*Corresponding authors

E-mail: aleksandr.egunov@main.tu-chemnitz.de,

E-mail: minshen.zhu@main.tu-chemnitz.de,

E-mail: daniil.karnaushenko@main.tu-chemnitz.de,

E-mail: oliver.schmidt@main.tu-chemnitz.de.

## Highlights

- A bottom-up, strain-induced self-assembly strategy transforms 2D thin-film stacks into 3D coaxial Swiss-roll microsystems.
- Monolithic integration of a lithographically patterned, chemically tunable polyimide nanomembrane serves as a programmable proton-exchange component.
- Ultra-compact bioelectronic power supply achieves a volumetric power density of ~3.1 mW $cm^{-3}$ within a 0.80 µL active volume and a 4.16 $mm^2$ footprint.
- Sustainable dual-mode operation decouples microbial metabolism from power generation, eliminating biofouling and enhancing long-term stability.
- The platform demonstrates scalable fabrication (>85% yield) and provides a versatile architecture for integrable bio-hybrid devices.

### TOC

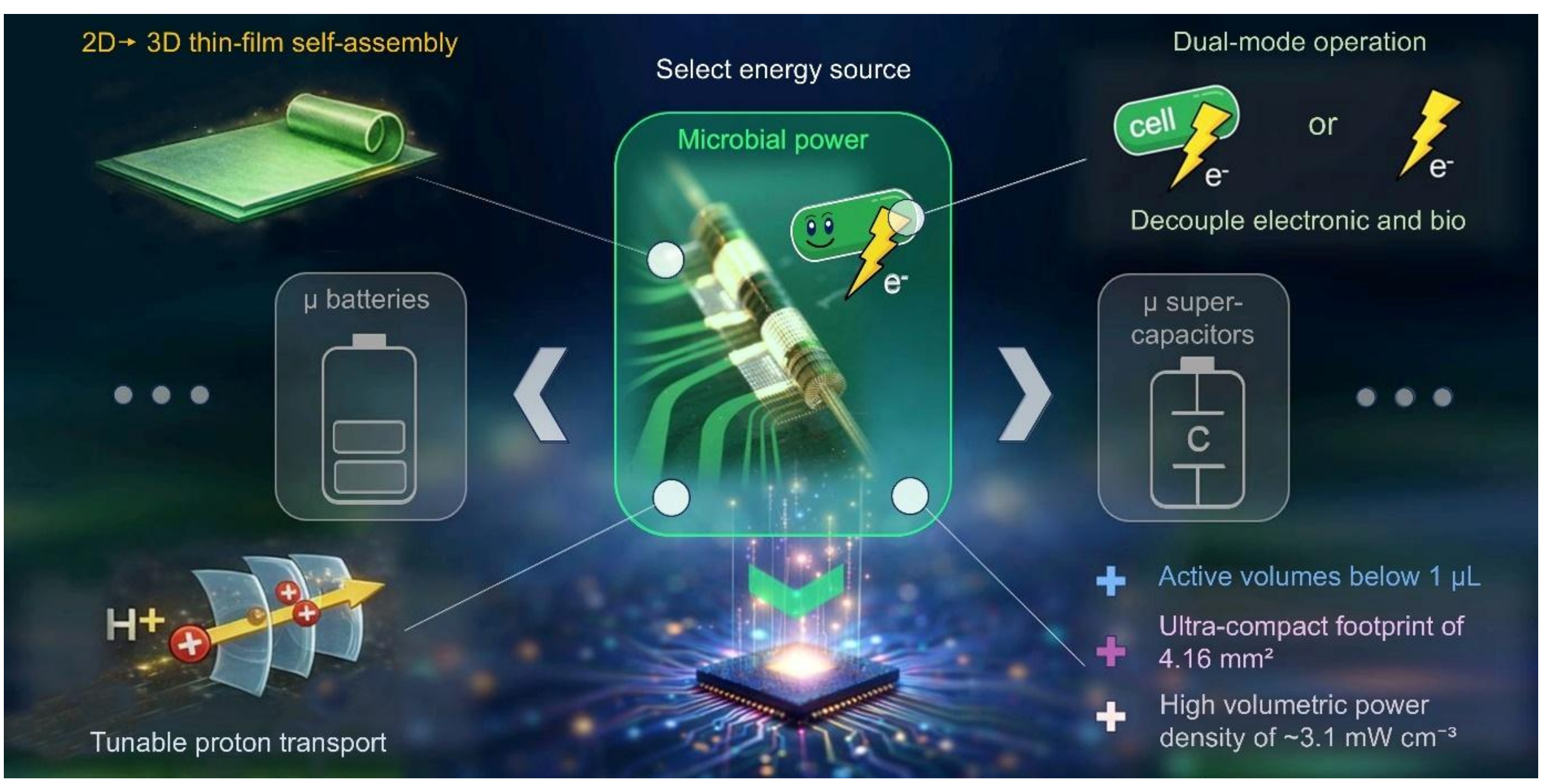

## Abstract

Thin-film self-assembly of three-dimensional (3D) microsystems presents a compelling route to integrate complex functionalities into ultra-compact volumes, yet strategies for incorporating tunable ion-conducting elements remain limited. Here, we introduce a strain-induced self-assembly platform that transforms lithographically patterned multilayer thin films into functional 3D coaxial Swiss-roll microtubes with total active volumes below 1 μL. A key innovation is the monolithic integration of a chemically tunable polyimide proton-exchange membrane, enabling post-fabrication optimization of ionic transport that balances proton transport with mediator blocking. We further implement a dual-mode operational scheme that decouples microbial metabolism from electrochemical power generation, revealing biofouling, not chemical fouling or membrane degradation, as the dominant failure mechanism in conventional architectures. Critically, optimally treated polyimide membranes exhibit excellent recoverability after fouling, while cell-free mode operation maintains stable performance by physically excluding microorganisms from the microelectronic environment. This integrated bio-electronic microsystem achieves a volumetric power density of ~3.1 mW $cm^{-3}$ within an ultra-compact footprint of 4.16 $mm^2$. Our work establishes a scalable thin-film engineering approach to create tunable, 3D bioelectronic power sources for autonomous microsystems.

**Keywords:** Self-assembly, 3D microfluidics, Thin-film membranes, Coaxial structure, Microbial fuel cells, Bioelectronics

## 1 Introduction

The development of autonomous microsystems and bioelectronic interfaces is fundamentally constrained by the lack of scalable, integrable ultra-small microsystems [1] and sensors [2,3]. To address this, next-generation microbial fuel cells (MFCs) must not only miniaturize but also adopt manufacturing processes compatible with microelectronics [4,5]. MFCs generate electrical power through microbial oxidation of organic substrates under mild conditions, offering sustainable energy conversion from inexpensive, renewable fuels [6].

Conventional devices that rely on direct electron transfer [7] between microorganisms and electrodes face critical limitations. Microbial biomass occupies a significant portion of the electrode volume, promoting the formation of inactive biofilm layers that increase internal resistance over time. In contrast, electron transfer mediated by soluble redox mediators eliminates the need for direct microbe–electrode contact while maintaining efficient redox cycling [8]. This strategy enables a new class of systems, termed indirect flow-type MFCs (IFT-MFCs), demonstrated here for the first time.

The IFT-MFC concept decouples microbial metabolism from electrochemical power generation. During operation, microorganisms such as *Saccharomyces cerevisiae* reduce soluble mediators via bio-oxidation in an external bioreactor. The reduced mediators are then delivered to the anode chamber, where they transfer electrons to the electrode surface. On the cathodic side, an oxidized catholyte (e.g., ferri/ferrocyanide) is regenerated by atmospheric oxygen and recirculated. This modular design enables dual-mode operation: a cell-based mode (Mode 1) for continuous mediator regeneration and a cell-free mode (Mode 2) for abiotic power generation, enhancing robustness and control.

At the microscale, microfluidic architectures provide a natural platform for implementing this concept [9], offering precise management of fluids and reactants [10] and improved kinetics [11]. The proton exchange membrane (PEM) plays a crucial role by facilitating ionic transport while preventing electrolyte mixing. Although hydrodynamic systems relying on laminar flow without a membrane have been explored, they require millimeter-scale electrode separations that increase internal resistance and offset the benefits of miniaturization [12]. Maintaining strict flow separation is essential, as even minor crossover can disrupt anaerobic conditions and cause electrochemical short-circuiting [13]. While laminar-flow systems [14] reduce mixing, membrane-based architectures [15] remain the most reliable for preventing self-discharge and ensuring long-term stability [16]. However, widely used ion-exchange materials such as Nafion™ remain costly and difficult to integrate at microscale [17], motivating the exploration of alternative polymers [17,18].

At the microscale, conventional PEMs (Nafion™, cellulose acetate, polyvinylidene fluoride (PVDF)) pose additional challenges, including high interfacial resistance and poor compatibility with wafer-scale microfabrication [19–22]. To address these issues, polyimide (PI) has emerged as a mechanically robust, biocompatible, and microfabrication-compatible alternative [23]. PI is widely employed in microelectronics and enables ultrathin, scalable devices [24–27], as demonstrated in neural implants [28], implantable antenna [29] and sensors for biological applications [30,31]. Its flexibility also allows for 3D integration, increasing active surface area while maintaining a compact footprint. To fully realize these 3D integration benefits, a fabrication strategy capable of shaping PI into complex, high-surface-area microfluidic architectures is required.

To achieve greater miniaturization, microfluidic MFCs (μMFCs) have been developed using planar, layered architectures [16,32,33]. While effective, these designs are inherently limited in surface-area-to-volume ratio and by the integration challenges of conventional proton-exchange membranes. The strain-driven self-assembly of PI into 3D architectures offers a compelling alternative, yet this powerful geometric transformation has not been fully leveraged to create monolithic, membrane-based bio-electrochemical microsystems. Translating this concept to μMFCs requires co-integrating the ion-conducting membrane and independent fluidic networks within the self-assembled 3D geometry - a critical challenge that remains unaddressed.

Recent progress in self-rolled polymeric nanomembranes [29,34,35] and rolled-up microfluidics [36] has opened a new route toward highly integrated 3D energy systems. This strain-driven self-assembly approach represents a powerful bottom-up strategy for microsystems fabrication, where 2D nano- and micro-scale thin films are engineered to autonomously form complex 3D architectures. It combines miniaturization with structural and functional integration. When released, thin polymeric nanomembranes spontaneously form tubular microfluidic structures [30,37] that can be seamlessly integrated with on-chip electronic components such as impedance sensors [31] and thin-film transistors [28,38]. This coaxial geometry enhances ion transport and boosts energy density, as shown by recent demonstrations of tubular capacitors [39,40], supercapacitors [41], and batteries [42,43] highlight the versatility of such architectures for powering microscale systems.

In this work, we present a 3D Swiss-roll indirect flow-type micro–microbial fuel cell (IFT-μMFC) designed to address this integration challenge (Fig. 1). A lithographically patterned PI [44] film incorporating planar electrodes is transformed into a coaxial 3D architecture through a strain-driven self-rolling [45]. This self-assembly provides concentric microfluidic channels separated by PI walls that function as the PEM. The resulting device (Fig. 1a) comprises an inner catholyte channel and an outer multi-winding anolyte channel, with electrochemical power generation occurring within the self-rolled structure. Both channels are connected to external reservoirs that continuously supply fresh electrolyte via a microfluidic distribution network, enabling independent control of anolyte and catholyte composition and flow. The device's fabrication, microfluidic integration, and dual-mode operation are summarized in the roadmap and operating principle schematic (Fig. 1b,c).

This modular architecture provides exceptional flexibility: By adjusting the external fuel containers, the system can be scaled or reconfigured to match varying energy requirements of microsystems. The combination of 3D self-assembly and microfluidic modularity [46] enables independent optimization of microbial culture, mediator regeneration, and electrochemical conversion, thereby enhancing both energy output and operational stability. This integration establishes a versatile bio-electronic platform that monolithically combines a tunable ion-conducting membrane with high-surface-area 3D microelectrodes within a single, self-assembled structure. Overall, by supporting a dual-mode operational scheme, the Swiss-roll IFT-μMFC provides a scalable and functionally robust foundation for powering autonomous microsystems.

**a** Self-assembled coaxial Swiss-roll μMFC with two PPEs sections

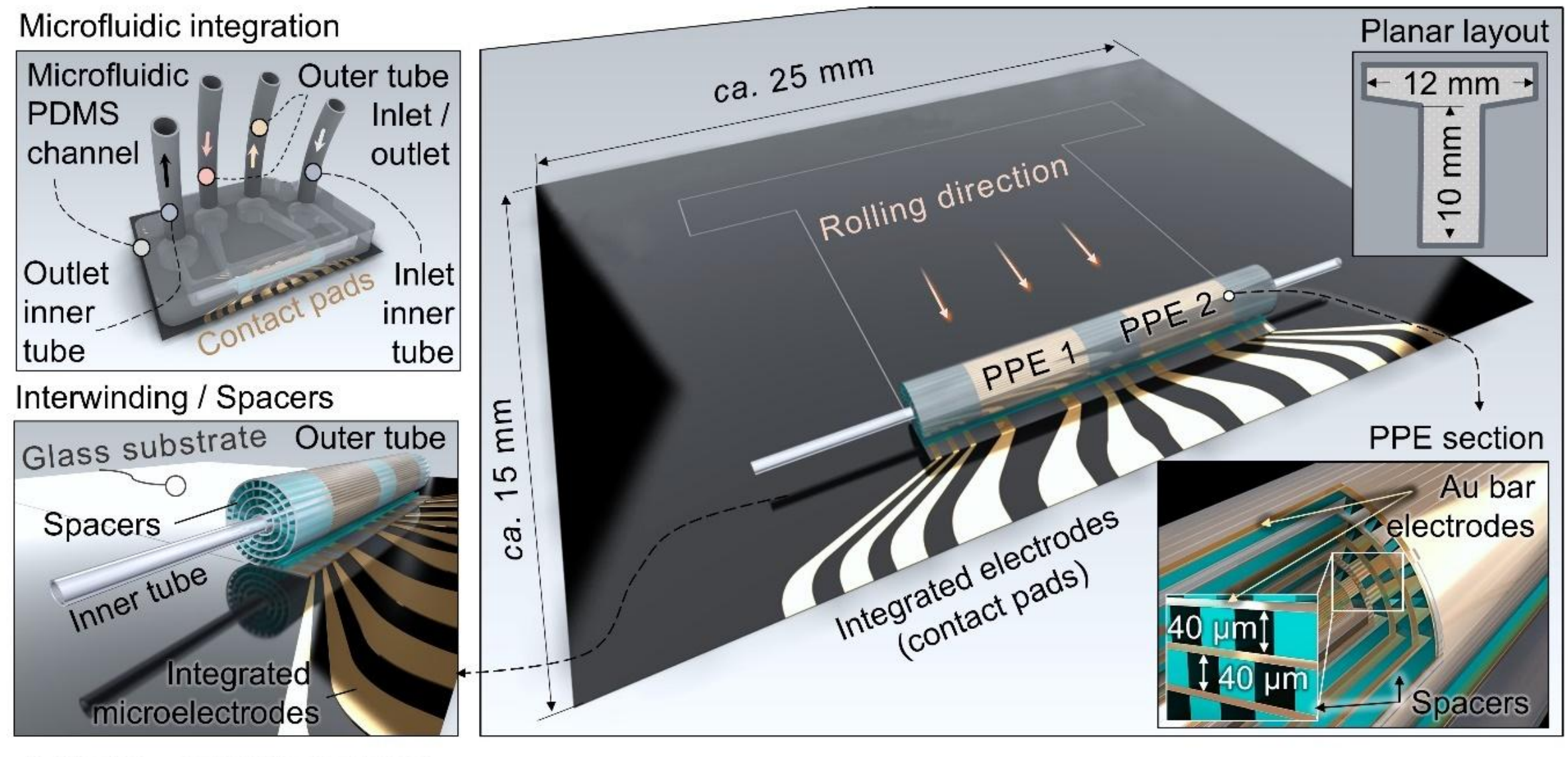

**b** Device assembly roadmap

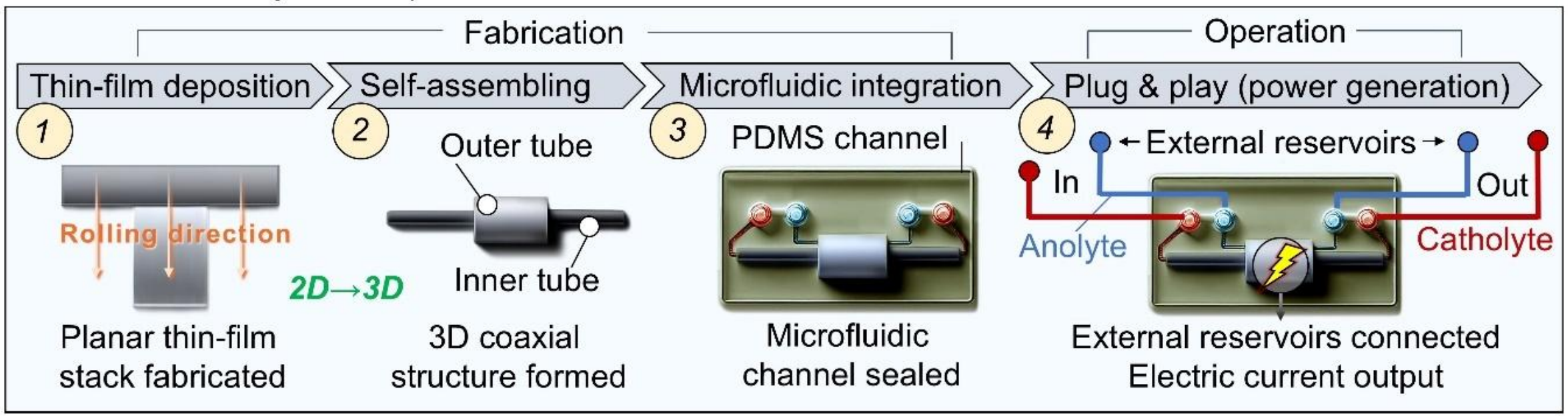

**c** Operating principles of coaxial Swiss-roll IFT-μMFC in dual mode

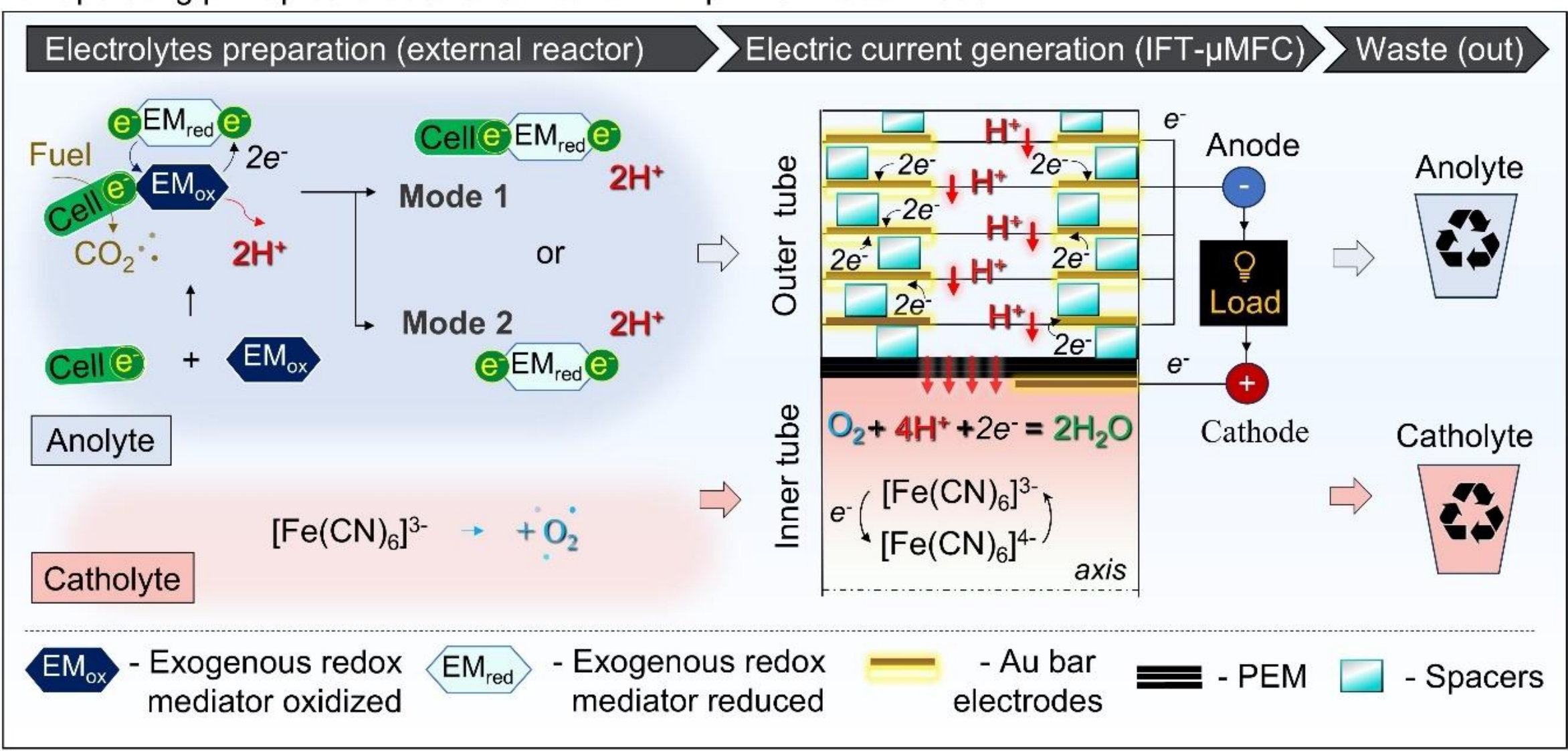

**Fig. 1** Concept and realization of the self-assembled coaxial Swiss-roll bio-electronic platform. **a** Schematic of the 3D self-assembled architecture integrated into a microfluidic system, showing the inner catholyte channel (inner tube) and the outer anolyte channel (outer tube) with integrated parallel-plate electrodes (PPEs). **b** Device assembly roadmap. Four key stages: (1) Thin-film deposition of the multilayer 2D stack. (2) Self-assembly via selective sacrificial layer dissolution and hydrogel swelling. (3) Microfluidic integration by alignment and sealing of the rolled tube within a PDMS chip. (4) Plug-and-play operation by connection to external fluidic lines and reservoirs for power generation. **c** Schematic illustration of the indirect flow-type operating principle, in which externally processed anolyte is continuously supplied under flow to the coaxial microfluidic device for electrochemical energy conversion. This flow-through architecture decouples the local electrochemical kinetics from the external regeneration or charging processes. The platform supports a dual-mode operational scheme for functional versatility: a microbial mode 1 (functioning as a pure microbial fuel cell with integrated biology) and a cell-free mode 2 (operating as an abiotic electrochemical generator using pre-charged mediator). This allows the system to dynamically alternate between regenerative bio-power and stable electrochemistry, providing a robust interface between biological and electronic subsystems.

## 2 Experimental section

### 2.1 Design and fabrication of self-assembled coaxial Swiss-roll microtubes

The devices were fabricated using standard thin-film techniques, building on strained-layer rolling principles. Briefly, a ~500 nm sacrificial layer (SL) was spin-coated from an aqueous solution of lanthanum(III) chloride heptahydrate (Thermo Scientific, 99%) containing 2-Benzyl-2-(dimethylamino)-4-morpholinobutyrophenone (DBMP, TCI, >98%) as a photoinitiator. A ~950 nm hydrogel (HG) layer was then spin-coated from a solution of 2-hydroxyethyl methacrylate (HEMA, Sigma-Aldrich, 97%), poly(ethylene-alt-maleic anhydride) (Sigma-Aldrich, Mw 100,000–500,000), and DBMP in N,N-dimethylacetamide (DMAc, Thermo Scientific, 99.5%). Subsequently, polyimide (PI) layers were deposited in sequence: first, an optional ~500 nm layer (used where stabilizing bars are required within the hydrogel), followed later by a ~1300 nm structural PI layer on top after intermediate processing steps, both spin-coated from a precursor solution containing 2-(dimethylamino)ethyl methacrylate (DMAEMA, Alfa Aesar, 97%), 4,4′-diaminodiphenylmethane (DDM, Sigma-Aldrich, ⩾97%), 3,3′,4,4′-benzophenonetetracarboxylic dianhydride (BTDA, Alfa Aesar, 97+%), and DBMP in DMAc, followed by thermal imidization. This material system for strain-engineered self-assembly builds upon established formulations detailed previously. Each layer was patterned by standard photolithography processes after deposition [29]. A reinforcing layer with a total thickness of ~1.3 μm was built up in the first winding region by the successive deposition and patterning of two (or three) thinner polyimide layers to ensure mechanical stability without compromising flexibility. SU-8 spacers of varying thicknesses were lithographically defined to control interwinding spacing and act as fluidic channels. Anchors were formed by patterning windows in the SL to prevent lateral rolling.

After fabrication, the structures were protected by a ~2.5 μm thick layer of AZ5214e photoresist. The samples were then treated with oxygen plasma (400 W, 10 min) to remove fabrication residues and defects along the structure edges. This plasma treatment simultaneously thinned the sacrificial layer from its original ~500 nm to approximately 150 nm, as the polymeric matrix of the SL is partially etched during the process. The photoresist was subsequently removed, revealing the final, defect-free structures with optimized rolling geometry. The self-rolling process was initiated by selective dissolution of the SL combined with HG swelling, forming an inner tube with interdigitated electrodes (IDEs) and an outer tube with parallel-plate electrodes (PPEs). To remove fabrication residues, the structures were treated with oxygen plasma (400 W, 20 min) through a ~1.4 μm AZ5214 photoresist mask. Rolling was initiated by immersing the released film stacks in a 0.5 M diethylenetriaminepentaacetic acid (DTPA, pH 8.0) solution at 22–24 °C for 24 hours**.** This process reproducibly yielded ~12 mm long coaxial tubes with diameters tunable from ~250 μm to >500 μm. Fabrication yield exceeded 85%.

Camphor-assisted drying: To preserve structure for microfluidic integration, coaxial rolls were immersed in a saturated camphor (≥98%, Alfa Aesar) solution in acetone. Upon withdrawal, a thin camphor layer crystallized as acetone evaporated. Complete sublimation under ambient conditions (4-5 h) left no residue and maintained winding uniformity.

Characterization: All deposition steps were monitored by stylus profilometry (Dektak, Bruker). Film quality and final structures were inspected by optical microscopy (Axioscope A1 and Keyence VHX, Zeiss). Scanning electron microscopy (SEM) was performed using a TESCAN GAIA3 microscope.

**2.2 Microfluidic integration of coaxial system**

Polydimethylsiloxane (PDMS) devices were fabricated by standard soft lithography using silicon wafer molds. PDMS (Sylgard 184, Dow, 10:1 base:curing agent) was cast, cured at 60 °C for 8 h, baked at 100 °C for 1 h, then demolded and port-punched.

Selective sealing: Uncured PDMS prepolymer was injected through six auxiliary lateral channels. The device was immediately placed on a 130 °C hotplate for 30 s to rapidly cure the PDMS and halt capillary-driven flow.

Fluidic connection: Devices were connected to a syringe pump (Pump 11 Pico Plus Elite, Harvard Apparatus) via PTFE tubing (OD 1/32, ID 0.30 mm; Darwin Microfluidics).

The integrity and isolation of the fluidic channels were characterized by flowing colored solutions. The inner and outer channels were perfused with aqueous solutions of an orange-fluorescent dye (Rhodamine B, ⩾95%, Sigma-Aldrich), and a green-fluorescent dye (Fluorescein sodium salt, Sigma-Aldrich), respectively. For interconnection tests, a red food colorant was used. In open-configuration filling experiments, deionized water and, for enhanced visualization, a commercial food colorant were used.

## 2.3 Electrical integrity and geometry-dependent electrode characterization

Electrodes (3D IDE, PPE, and IDE–PPE sets) were characterized in a four-probe configuration to minimize cable parasitics. EIS was performed using a high-precision LCR meter (LCR-8105G, GW Instek) over 20 Hz – 5 MHz with a 500 mV AC excitation signal. Electrolytes included KCl (≥99.5%, ITW Reagents, $10^{-4}$–1 M in 8 MΩ×cm DI water) and ethanol–water mixtures (using $C_2H_5OH$, ⩾90%, AnalytiChem, diluted with DI water). Samples were equilibrated at 22–24 °C for 30 min prior to testing. Impedance data were analyzed to extract real/imaginary components and apparent capacitance.

## 2.4 Tuning proton exchange in a multi-winding membrane architecture

Lithographically fabricated PI PEMs, with a final thickness of ~1.3 μm, were mounted in PDMS frames. Proton transfer was evaluated in a macro-scale setup with two 3D-printed acrylic chambers (6 mL each), by adding 1 mL of 1 M HCl to 5 mL of 1 M KCl in the anodic chamber and monitoring the cathodic chamber pH over time. Membrane thickness was varied from 1 to 4 layers.

For resistance tuning, membranes were immersed in aqueous NaOH (1M, Sigma-Aldrich, ≥98%) at 22 °C for 1-60 min, achieving resistances from ~9 kΩ (1-2 min) to ~250Ω (~30 min). Prolonged immersion (>1 h) reduced resistance below 100Ω but induced fragilization.

PEM resistance ($R_{mem}$) extraction: $R_{mem}$ was determined from EIS in symmetric two-electrode cells using parallel Au electrodes (area 1300 $mm^2$ each) in 1 M KCl. Resistance was extracted as the real component at $10^3$ Hz. Control experiments used Nafion™ 117 (DuPont, 183 µm) and Kapton™ (DuPont, 125 µm). Each condition was tested in triplicate.

FT-IR characterization: FT-IR spectra were acquired using a Nicolet iS50 spectrometer (Thermo Scientific), averaging 128 scans at 4 $cm^{-1}$ resolution under nitrogen purge.

**2.5 Component-level optimization of a bio-electrochemical energy system**

*Saccharomyces cerevisiae* (baker's yeast, BY4741) was cultured anaerobically in YPD medium with 2% glucose (alpha-D(+)-Glucose, ≥99%, Thermo Scientific) and introduced into the anodic chamber under anaerobic conditions.

Methylene blue (≥98%, Thermo Scientific) was used as the anodic exogenous mediator (EM) at concentrations from $1.25 \times 10^{-6}$ M to 0.1 M in PBS (Pan Biotech, pH 7.4). The catholyte consisted of potassium ferricyanide (≥99%, Sigma-Aldrich) dissolved in a 1:1 mixture of 1 M KCl and glacial acetic acid. The EM concentration was monitored using an LLG-uniSPEC 1 UV/VIS spectrophotometer with single-use plastic cuvettes. Control measurements were carried out with an Infinite M Nano+ UV/VIS spectrophotometer microplate reader (Tecan) at the characteristic absorption maximum of methylene blue (660 nm). Single-use plastic plates Greiner 96 Flat Bottom Transparent Polystyrene were used throughout. A calibration curve was constructed from standard solutions in 1M KCl covering the relevant concentration range from 0.05 M down to approximately 50 nM ($10^{-8}$-$10^{-1}$ M). Prior to each measurement session, a background spectrum of KCl was subtracted. All measurements were performed using 4 reads per well. For each well, the mean absorbance was calculated, and the deviation (standard deviation) among the four reads was used to represent the measurement variability in the plotted data.

Macro-scale MFCs were fitted with Au electrodes (10 nm Cr / 100 nm Au, 50–600 $mm^2$) or graphite rods (99.95%, Sigma-Aldrich, 32–475 $mm^2$). Polarization and power density curves were obtained by linear sweep voltammetry (LSV, 1 mV $s^{-1}$, Autolab PGSTAT101) from open-circuit voltage (OCV) to 0 V.

Biofouling analysis: PI PEM and gold electrode samples were immersed for 24 h in sterile KCl (control), 0.05 M MB solution (organic fouling), or complete microbial anolyte (biofouling). After immersion, samples were rinsed with PBS, fixed with 4% Formaldehyde (12h, 4°C), and dehydrated through a graded ethanol series with subsequent Hexamethyldisilazane (HMDS, Sigma-Aldrich) drying. Following the drying, samples were sputter-coated with 5 nm Cr and imaged by SEM (TESCAN GAIA3 microscope).

### 2.6 Sustainable dual-mode operation and performance benchmarking of a coaxial Swiss-roll bio-electronic platform

For cell-free operation, the anolyte was anaerobically filtered through 0.22 μm PVDF membranes (Millipore) to remove yeast cells, yielding a purified solution of reduced mediator (EMred).

Coaxial μMFCs employed PI membranes treated in NaOH (1 M, 30–180 min, 50 °C), achieving resistances from ~100 Ω to >1 MΩ. Polarization and power curves were recorded as described above.

Long-term stability tests were performed for 24 h at room temperature using the optimized anolyte and catholyte compositions described above, with continuous recirculation at 0.1 rpm (Gilson Minipuls 3 peristaltic pump, 10-roller pump head) using platinum-cured silicone tubing (0.88 mm ID × 1.6 mm wall thickness, Darwin Microfluidics), corresponding to a flow rate of approximately 12 μL/min. Microbial culture was maintained in the external bioreactor throughout the microbial mode operation.

### 2.7 Calculation of Active Volume and Volumetric Power Density

The active volume of the coaxial Swiss-roll structure refers strictly to the internal fluidic volume where electrochemical reactions occur. It was calculated geometrically from the known dimensions of all lithographically defined elements. This volume was determined individually for both the inner catholyte channel and the outer anolyte channel, excluding all external tubing, connectors, and reservoirs.

### 2.8 Statistical Analysis and Data Presentation

All quantitative experiments were performed on at least three independently fabricated devices ($n \geqslant 3$) unless otherwise specified. Results are reported as mean $\pm$ standard deviation (SD).

Electrochemical impedance spectroscopy (EIS) measurements were performed with an instrumental accuracy of ±1% in impedance magnitude (LCR-8105G, GW Instek). Each reported spectrum represents the average of four consecutive sweeps per device, and measurements were reproduced on at least three independent devices to verify reproducibility. Due to the high frequency resolution (200 points per spectrum), error bars at individual frequencies are not displayed. Device-to-device variability in impedance magnitude remained below 8% across the tested frequency range.

Layer thicknesses were verified by stylus profilometry (Dektak XT, Bruker) on test substrates processed alongside device wafers. For each layer, at least 10 measurements were collected. For readability, nominal thickness values are reported in the text (e.g., ~250 nm, ~900 nm, ~1.3 µm); measured values were within ±5-6% of these nominal targets with coefficients of variation below 5%.

Polarization and power density curves represent mean values obtained from independent devices. For datasets where device-to-device variability exceeded 5% of the mean value, SD error bars are displayed directly on the figures at representative current densities. For plots where error bars are not explicitly shown, the relative SD was below 5% of the mean value for the representative data points tested.

Spectrophotometric calibration curves were generated from at least four standard concentrations measured in triplicate, with linear regression coefficients ($R^2$) exceeding 0.999.

Differences were considered statistically significant at $p < 0.05$. Statistical analyses were performed using OriginPro (version 2020).

## 3 Results and discussion

### 3.1 Design and fabrication of self-assembled coaxial Swiss-roll microtubes

We fabricated a polymeric nanomembrane stack using a strain-engineered self-rolling approach, building on previously reported methods [28,29]. This bottom-up self-assembly strategy transforms a lithographically patterned 2D thin-film stack into a functional 3D microsystem. The multilayer design integrated a sacrificial layer (SL), polyimide (PI), a hydrogel (HG) layer, thin-film gold electrodes, and SU-8 spacers, enabling controlled self-assembly and the direct integration of electrodes within the membrane stack (see details in Experimental section, Supplementary Figs. S1, S2).

Upon selective dissolution of the sacrificial layer, a combination of built-in stress gradients and hydrogel swelling reshaped the planar stack into coaxial Swiss-roll tubes with precisely defined winding geometries. The design of the two-dimensional precursor was optimized to ensure stable rolling: a T-shaped layout proved optimal for the sequential formation of a stable coaxial configuration comprising an inner long tube and an outer short tube (Fig. 2 and Supplementary Figs. S2, S3). The HG layer was lithographically patterned into three primary stripes beneath the electrode regions, with continuous sections retained at the structure's beginning (inner tube) and structure's ends (adhesion part) (Supplementary Fig. S1). This design provides sufficient strain gradient for controlled rolling while ensuring the electrode-bearing PI membrane remains largely free of hydrogel, defining the primary proton transport interface.

Optical microscopy confirmed the reproducible formation of ~12 mm-long coaxial structures with tightly aligned windings. The SU-8 spacer bars defined interwinding gaps from ~5 µm to 50 µm, providing direct control over the final tube diameter, which ranged from below 250 µm to above 500 µm. Scanning electron microscopy revealed mechanically robust rolling with no delamination or interfacial defects, confirming the structural integrity of the self-assembled architecture (Supplementary Fig. S4). The transition between the inner and outer tubes was stabilized by a ~1.3 µm PI reinforcing layer in the first winding region, which prevented delamination and ensured a smooth curvature. Furthermore, lithographically defined anchors, directly coupled to the glass substrate, prevented uncontrolled lateral rolling and secured consistent coaxial assembly across the wafer.

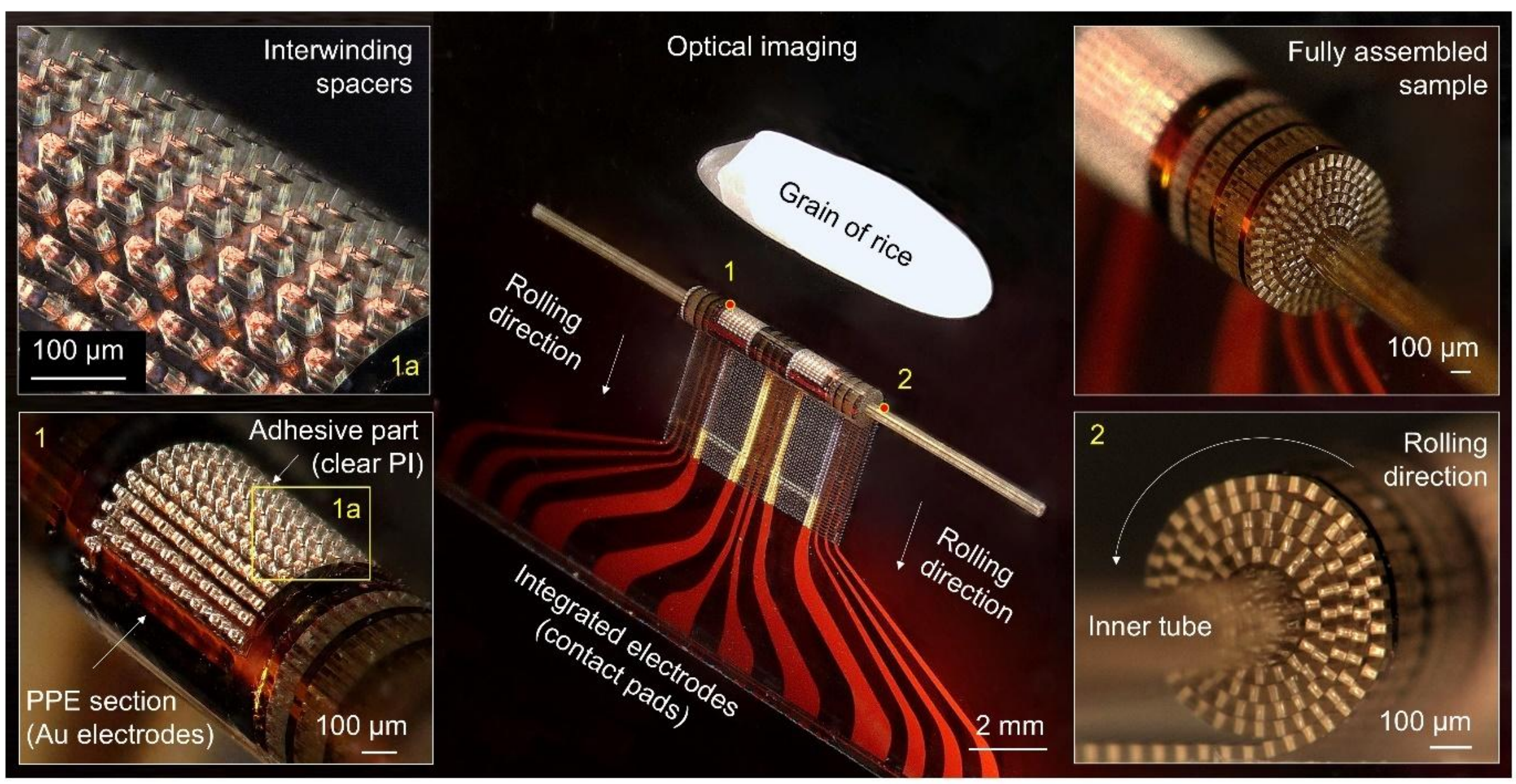

**Fig. 2** Structural characterization of the self-assembled coaxial Swiss-roll microtubes. Optical microscope images of the fabricated 3D architecture after release from the substrate. The main panel shows an overview of the structure, highlighting the simultaneous formation of the inner and outer tubes. The **left** panel provides a magnified view of the SU-8 spacer bars that define the interwinding gaps and fluidic channels. The **right** panel shows a higher-magnification side view of the tube entrance, revealing the tightly packed and uniform windings that ensure mechanical stability.

The fabricated devices exhibited a yield exceeding 85% for mechanically intact coaxial structures per wafer, with minimal dimensional variability. These results demonstrate that the self-rolled fabrication strategy enables the reproducible integration of electrodes, PI membranes, and spacers into mechanically robust 3D architectures, representing a prime example of bottom-up microsystems engineering. By tuning spacer dimensions and reinforcement layers, the tube diameter and interwinding geometry can be precisely engineered, establishing a structurally versatile and scalable platform for bio-electronic microsystems.

### 3.2 Microfluidic integration of coaxial system

The self-assembled coaxial structures were integrated into a polydimethylsiloxane (PDMS) microfluidic platform, enabling independent and stable fluidic operation of both the inner and outer channels (Fig. 3a). Prior to integration, the rolled structures were carefully dried using an established protocol [30] that exploits camphor sublimation to gradually replace residual water, thereby minimizing capillary forces and structural stress while preserving the intact coaxial geometry (see Experimental section).

Precise alignment of the coaxial tubes within the PDMS mould under optical microscopy ensured correct port registration. Liquid PDMS, delivered through six auxiliary lateral channels, selectively encapsulated the device while preserving the open interwinding channels, yielding reproducible sealing without structural collapse (Supplementary Fig. S5a). The procedure consistently produced mechanically intact and leak-free sealing across samples, without collapse of the tubular geometry. Photographs of the integrated chip confirmed robust interconnection with PTFE tubing and a compact footprint compatible with on-chip integration (Fig. 3a, right). Crucially, after self-assembly and camphor-assisted drying, the coaxial structure is permanently encapsulated and mechanically fixed within the PDMS chip. This prevents any subsequent dimensional changes in the hydrogel, decoupling its initial mechanical role from the operational stability of the microfluidic and electrochemical system.

The device architecture allows for independent injection of fluid streams through the coaxial channels while maintaining complete isolation between compartments, as illustrated in the connection scheme (Fig. 3b). The integrity of the microfluidic architecture was confirmed through sequential filling experiments. These tests demonstrated uniform wetting of both coaxial tubes and a controlled, sequential filling of the interwinding channels (Supplementary Fig. S5b and Supplementary Videos V1–5). This progression, guided by the SU-8 spacer design, confirms efficient fluidic access to the entire 3D geometry and demonstrates the potential for localized fluidic addressing within the coiled architecture. The stable microfluidic operation under bidirectional flow confirms the mechanical integrity of the encapsulated Swiss-roll structure, with no observable winding looseness or delamination.

This robust and reproducible integration strategy combines self-assembled coaxial nanomembrane tubes with soft lithographic microfluidics, enabling precise dual-channel fluid control within a fully sealed microsystem. The resulting continuous spiral channel geometry, devoid of dead volumes or junctions, is inherently favorable for fluidic operations as it promotes the advection of potential gaseous byproducts (e.g., $CO_2$ from microbial metabolism) and prevents their local accumulation, mitigating risks associated with flow blockage.

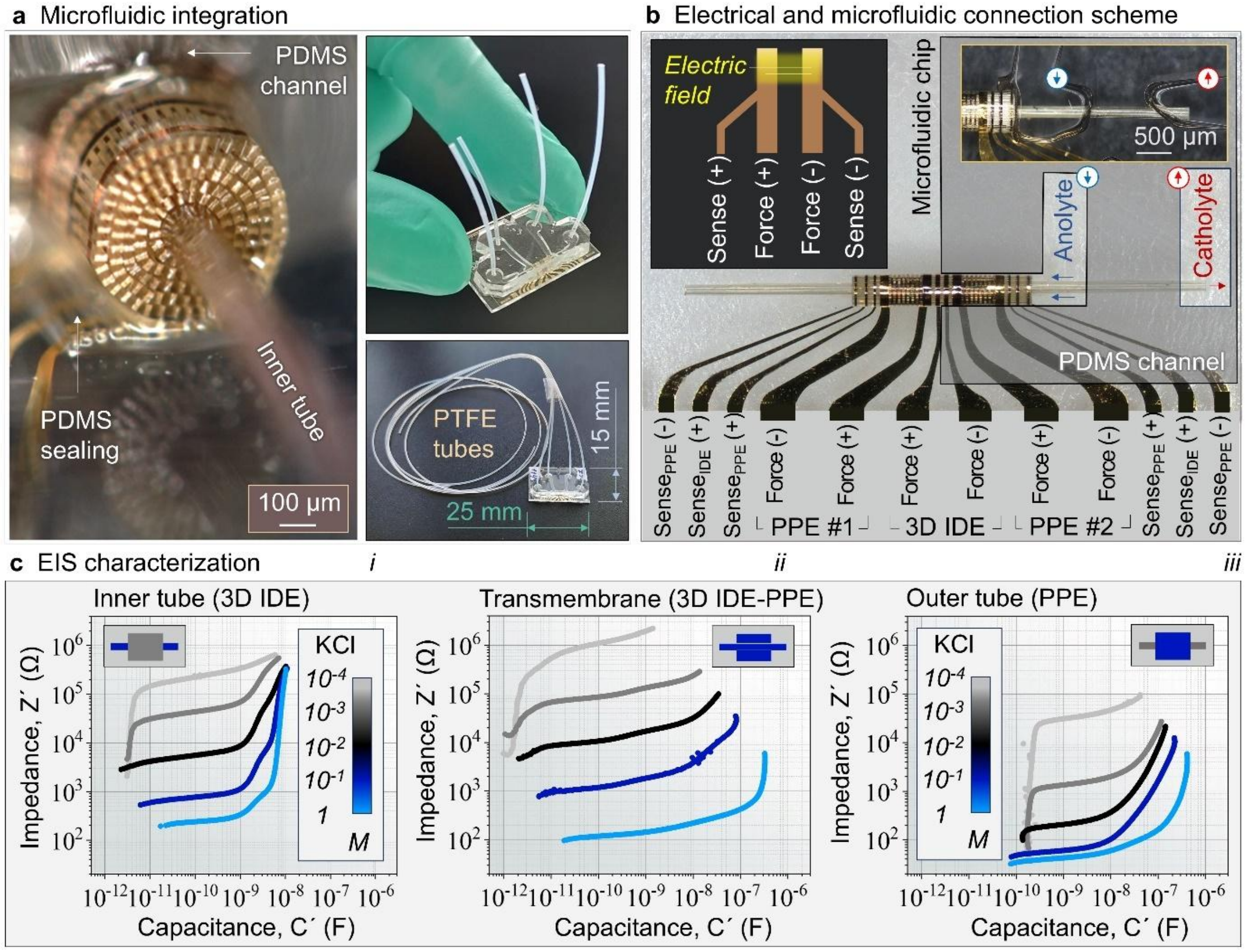

**Fig. 3** Microfluidic integration and electrical characterization of the coaxial platform. **a** Photographs of the device: the left image shows an open configuration with selective sealing of the inner and outer tubes, illustrating the preserved interwinding channels; the right image shows the fully assembled PDMS microfluidic chip with fluidic interconnects, **b** Electrical and microfluidic connection scheme, detailing the independent supply of fluids to the inner (catholyte) and outer (anolyte) channels, **c** Electrochemical impedance spectroscopy (EIS) performance of the integrated electrodes across a range of KCl solutions ($10^{-4}$ to 1 M). The data for the 3D interdigitated electrodes ( 3D IDEs) (*i*), the 3D IDE–parallel-plate electrode (PPE) pair (*ii*), and the PPEs (*iii*) demonstrate that PPEs exhibit the lowest impedance, while the 3D IDEs provide stable performance across ionic strengths.

### 3.3 Electrical integrity and geometry-dependent electrode characterization

Electrochemical impedance spectroscopy (EIS) characterization of the integrated platform confirmed its electrical robustness, revealing low interfacial resistance and stable electrode connectivity (Fig. 3c). This validates that the self-assembly and microfluidic integration processes preserve the electrical integrity required for precise readout. To evaluate performance, EIS was conducted across KCl concentrations from $10^{-5}$ to 1 M (Fig. 3c, Supplementary Figs. S6–S7; Supplementary Information, Experimental Section). Capacitance–impedance spectra (C′ vs. Z′, Fig. 3c) revealed distinct responses for each electrode geometry. At low ionic strength, all electrodes showed reduced capacitance, consistent with the increased thickness of the electrical double layer's diffuse region. This effect was most pronounced for planar 2D interdigitated electrodes (IDEs). Increasing ionic strength decreased the double-layer thickness, thereby increasing the measured capacitance and decreasing the impedance. Increasing ionic strength stabilized capacitance and decreased impedance. Parallel-plate electrodes (PPEs) exhibited the most extended capacitive plateau, consistent with their larger effective surface area, whereas 3D coaxial IDEs provided a broader frequency window of stable double-layer charging. This trend is quantitatively explained by the contraction of the electrical double layer with increasing ionic strength. The Debye screening length ($\lambda_D$) decreases from ~100 nm at $10^{-5}$ M KCl to ~0.3 nm at 1 M KCl (Supplementary Table S1), leading to more efficient double-layer charging and a corresponding order-of-magnitude decrease in impedance.

The corresponding Bode plots (|Z| vs. frequency; Supplementary Fig. S7) supported these findings. 3D IDEs maintained lower impedance across the full frequency range, while PPEs

exhibited suppressed impedance at intermediate to high frequencies ($10^3$–$10^5$ Hz), indicating enhanced charge-transfer characteristics. To probe sensitivity to dielectric environments, measurements were performed in ethanol–water mixtures (Supplementary Fig. S8). Decreasing the solvent dielectric constant increased impedance and reduced capacitance across all geometries. PPEs were most sensitive, showing nearly three orders of magnitude change in impedance at 100 Hz, while 2D IDEs remained stable and 3D IDEs displayed intermediate behaviour. This dielectric dependence aligns with double-layer theory and highlights how geometry can tailor environmental sensitivity.

Collectively, these results confirm the electrical robustness of the platform and demonstrate that electrode geometry dictates the electrochemical response. PPEs maximize interfacial area and dielectric sensitivity, while 3D IDEs provide stable charge transfer across frequencies, establishing a versatile foundation for integrated microscale electrochemical systems. PPE geometry, which exhibited the lowest impedance and most stable capacitive response across the relevant ionic strength range (0.1–1 M), was selected for integration into the Swiss-roll μMFC. This choice directly minimizes ohmic losses within the device and contributes to the high power densities achieved.

### 3.4 Tuning proton exchange in a multi-winding membrane architecture

The proton exchange membrane (PEM) governs ionic transport in the coaxial Swiss-roll platform and thus directly determines device performance. Fig. 4a shows a schematic cross-section of the architecture, highlighting proton transport pathways across a single polyimide (PI) membrane layer between adjacent windings, and across multiple layers between the inner and outer tubes. This multi-winding geometry creates distinct ionic conduction paths and underscores the role of membrane properties in electrochemical behaviour. SEM (Fig. 4b) resolved the winding morphology and spacer architecture, confirming uniform interwinding spacing and mechanically stable channels that act as controlled ion-conducting pathways.

Proton transfer dynamics were evaluated by introducing HCl and monitoring pH variation across membranes over time (Fig. 4c). The ~1.3 μm thick PI membranes were systematically treated in NaOH, achieving resistances from >1 MΩ (pristine) down to ~250 Ω after optimal conditioning. This tunability originates from a controlled hydrolysis process (Fig. 4d). The mild alkali treatment partially opens imide rings (Fig. 4d, i), introducing carboxylate and amide groups that increase hydrophilicity and create hydrated proton-conducting domains, as confirmed by FTIR spectroscopy (Supplementary Fig. S9a), thereby improving proton mobility. However, excessive hydrolysis disrupts polymer chain cohesion, leading to swelling and mechanical fragilization. This trade-off defines an optimal processing window for balancing conductivity and stability.

Based on this trade-off, moderate treatments (yielding ~9 kΩ–500 Ω) (Fig. 4d, *ii* and Supplementary Fig. S9b) significantly improved proton transfer relative to untreated PI, while the most efficient exchange was achieved at ~250 Ω. When compared to commercial PEMs (Fig. 4d, *iii*), a single optimally treated PI layer approached the performance of Nafion™ 117, while additional layers slowed transport yet still outperformed Kapton™. The balance of conductivity and mechanical robustness led us to adopt ~250 Ω PI membranes for subsequent device integration.

To assess dimensional stability, we quantified the swelling behavior of pristine and NaOH-treated PI membranes by optical tracking of lithographed metal markers (Supplementary Fig. S9c). Over 24 h immersion, both membranes exhibited minimal swelling (<2.5%), confirming that controlled hydrolysis (⩾~250 Ω) enhances proton conductivity without compromising structural integrity.

These results demonstrate that (i) proton exchange in multilayer Swiss-roll geometries can be finely tuned by the controllable treatment of individual ultrathin PI films, (ii) the platform inherently benefits from minimal local transport distances along winding interfaces, and (iii) conductivity can be modulated over several orders of magnitude. This pre-fabrication control over a fundamental device property establishes a versatile and scalable pathway for creating bio-electronic microsystems with customized ionic pathways.

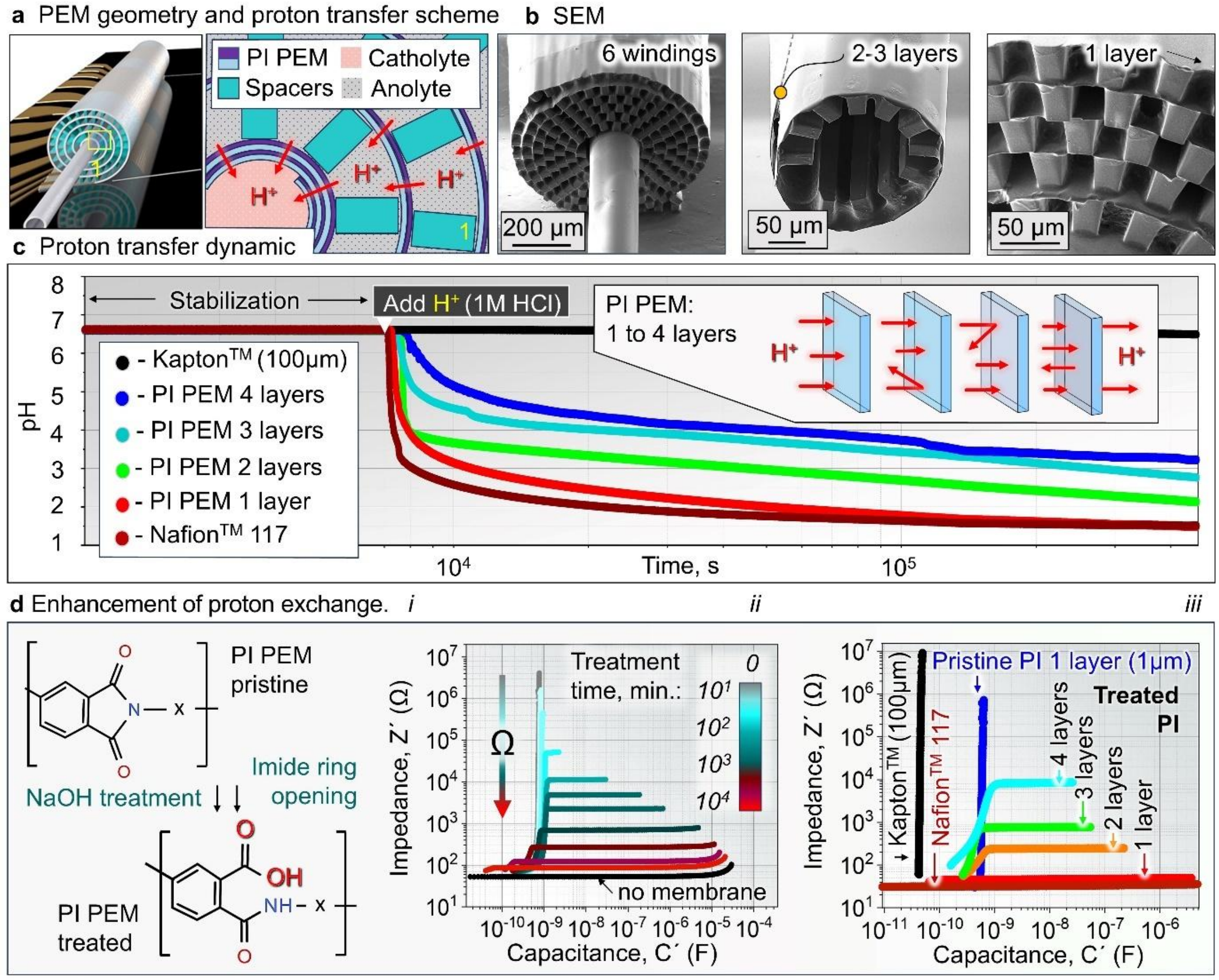

**Fig. 4** Enhanced proton transport via chemically tuned polyimide membranes in the Swiss-roll architecture. **a** Schematic cross-section illustrating proton transport pathways across single and multiple polyimide proton exchange membrane (PI PEM) layers in the multi-winding geometry. **b** SEM resolving the winding morphology, showing regions with single and multiple polyimide layers (corresponding to the pathways in section a) and confirming the mechanically stable, uniform interwinding spacers. **c** Proton transfer dynamics, showing pH variation over time for treated PI PEMs of different layer counts (1-4 layers) compared to commercial benchmarks (Nafion™ 117, Kapton™) after $H^+$ addition. **d** Chemical tuning and performance. Mechanism of NaOH-induced hydrolysis, where partial imide ring-opening creates hydrophilic proton-conducting domains (*i*). Membrane resistance decreases with treatment time, defining an optimal window at ~250 Ω (*ii*). Benchmarking shows a single tuned PI layer approaches Nafion™ 117 performance, compared to multi-layer PI stacks and Kapton™ (*iii*).

### 3.5 Component-level optimization of a bio-electrochemical energy system

To identify the key parameters governing system efficiency, we established a macro-scale model where electrodes, the proton exchange membrane (PEM), and the redox mediator could be varied independently. *Saccharomyces cerevisiae* served as the microbial catalyst, oxidizing glucose and reducing the anodic exogenous mediator methylene blue, which subsequently transferred electrons to the anode. The catholyte contained a ferri/ferrocyanide redox couple as a stable electron acceptor.

Electrode geometry emerged as a primary determinant of performance. Increasing the anodic surface area consistently raised both current density and maximum power output (Fig. 5, *i–ii)*. Large anodes (>200 mm²) supported currents above 150 μA and peak power exceeding 20 μW. In contrast, enlarging the cathode provided only modest improvements. These results confirm that the anodic reaction, which is driven by mediator cycling between microbial metabolism and the electrode, is the main system bottleneck.

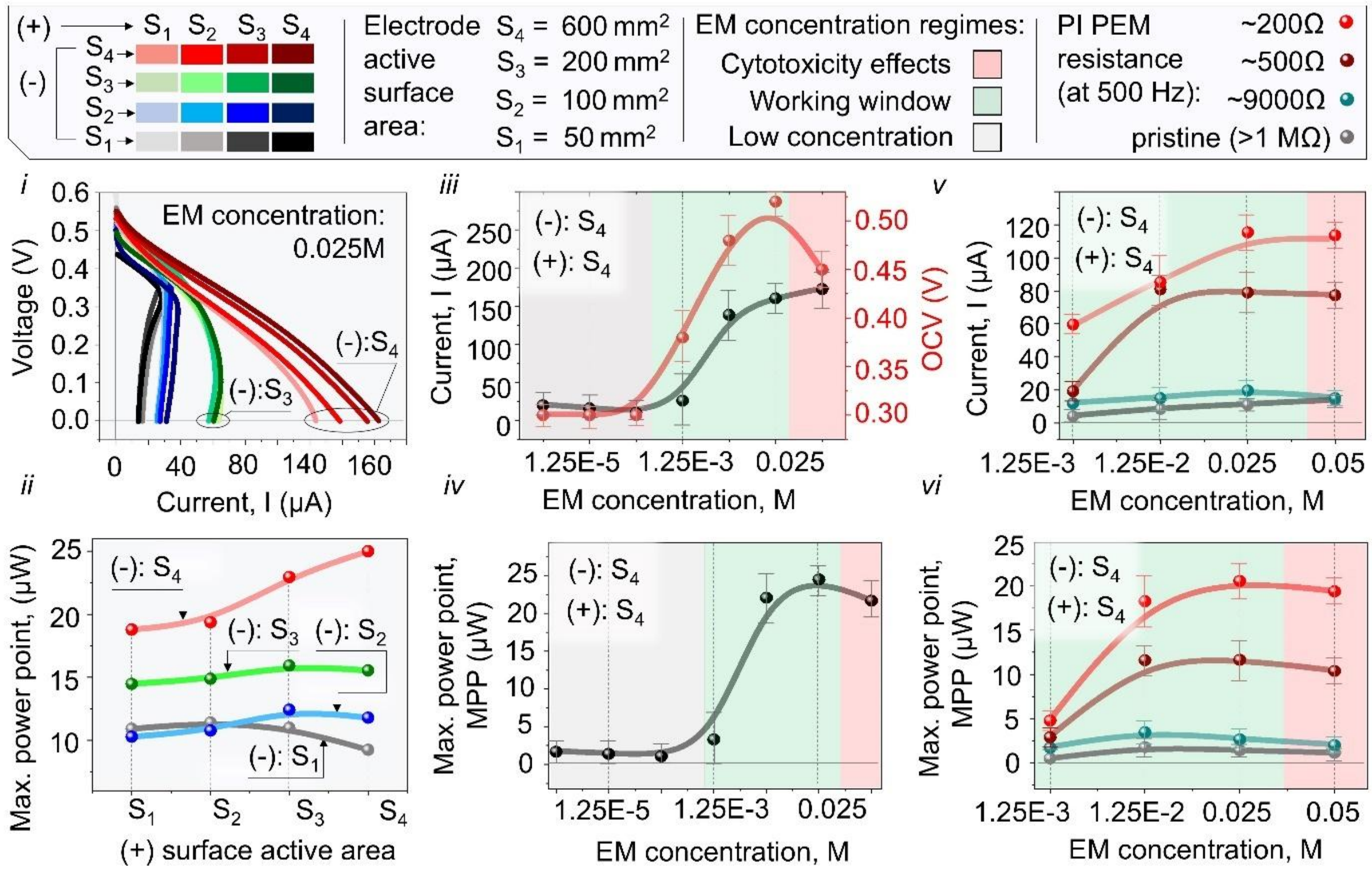

**Fig. 5** Component-level optimization of the bio-electrochemical system. Influence of electrode geometry (*i-ii*), exogenous mediator (EM) concentration (*iii-iv*), and proton exchange membrane (PEM) resistance on system performance (*v-vi*). increasing anode surface area enhanced current and power, with large anodes (>200 mm²) supporting >150 μA and >20 μW, while cathode enlargement yielded minimal gains (*i-ii*). Mediator concentration exhibited an optimal window (~0.0125–0.025 M), lower concentrations limited electron transfer while higher levels induced cytotoxicity (*iii-iv)*, PEM resistance was critical: low-resistance PI

membranes (~200 Ω) enabled high performance, whereas intermediate (~500 Ω) and high (⩾ 9 kΩ) resistance severely limited output (*v-vi*). These results establish key design rules: maximize anode area, maintain mediator within a biocompatible range, and minimize PEM resistance. Data points are measurements; lines are guides for the eye (*ii-vi*).

The concentration of the exogenous mediator (EM) defined a narrow operational window (Fig. 5, *iii–iv*; Supplementary Fig. S10). Optimal performance was observed in the $1.25\times10^{-3}$ to 0.025 M range, yielding power densities >20 μW at >150 μA. However, very low concentrations (<$1.25\times10^{-5}$ M) failed to sustain effective electron transfer, and higher concentrations (⩾0.05 M) reduced output, an effect attributed to cytotoxicity. This biphasic response highlights the critical need to balance electrochemical kinetics with biological compatibility.

The PEM exerted dominant control over overall performance (Fig. 5, *v–vi*; Supplementary Figs. S10-S11). Untreated polyimide (PI) membranes with high resistance (>1 MΩ) nearly suppressed current and power. Our tunable PI membranes, when treated to achieve low resistance (~100-250 Ω), supported the highest performance, while intermediate (~500 Ω) and high (⩾9 kΩ) resistance caused a marked decline. These results emphasise that PEM resistance is a critical limiting factor.

The optimal performance window for mediator concentration (~$1.25\times10^{-3}$ to 0.025 M) remained consistent regardless of PEM resistance (Supplementary Fig. S11, S12). However, within this concentration window, the absolute power output was critically governed by the membrane resistance. Low-resistance PEMs (~100–250 Ω) enabled high current and power, whereas increasing resistance progressively limited performance, confirming that proton transport, not mediator kinetics, was the dominant limiting factor for optimally formulated electrolytes.

Together, these macro-scale experiments establish the key design rules for our coaxial integration: maximize anodic surface area, maintain mediator concentration within a biocompatible but electrochemically effective window (~$1.25\times10^{-3}$-0.025 M), and minimize PEM resistance using moderately treated PI membranes (~250-300 Ω). These systematic, component-level insights provided a direct and universal quantitative framework for rational device design, which we applied to our coaxial Swiss-roll μMFC.

### 3.6 Sustainable dual-mode operation and performance benchmarking of a coaxial Swiss-roll bio-electronic platform

To demonstrate a path toward sustainable performance and circular operation, we implemented a dual-mode operational strategy for the coaxial Swiss-roll platform that integrates microbial metabolism with mediator cycling (Fig. 6a). This concept enables the potential recycling of active species to minimize waste and facilitate long-term function. The core principle, which is the decoupling of biological regeneration from electrochemical power generation, was experimentally validated under both microbial and cell-free conditions (Fig. 6b). In the microbial mode, electroactive yeast cells oxidized glucose in an external bioreactor, reducing the soluble mediator. The reduced mediator ($EM_{red}$) (Fig 6a) was then delivered to the µMFC anode to sustain power generation. In the cell-free mode, the anolyte was filtered, leaving a purified $EM_{red}$ solution that independently powered the device. Critically, the polarization curves of both modes nearly overlapped (Fig. 6b, *i–ii*), with peak power outputs of ~1.7 µW. This confirms that pre-reduced mediators alone can sustain device operation, successfully establishing the functional decoupling that is the fundamental prerequisite for a recyclable, closed-loop system.

We next evaluated the influence of PI PEM resistance on dual-mode performance. At low resistance (~100 Ω), the µMFC reached current densities of ~8 µA with power maxima around 2 µW (Fig. 6b, *iii–iv*). Increasing membrane resistance to ~500 Ω reduced output, while highly resistive membranes (~10 kΩ) limited current to ~1 µA with negligible power. Electrochemical impedance spectroscopy (EIS) confirmed these findings (Fig. 6b, *v–vi*), with treated PI membranes (~100–500Ω) showing low, stable impedance and highly resistive membranes exhibiting transport-limited behaviour.

To benchmark our platform against the state of the art, we constructed a Ragone chart (Fig. 6c) comparing its performance to other reported microbial fuel cells (MFCs) with membrane-based architectures and with active electrolyte volumes below 1 mL [16,32,33,47–50]. Our coaxial Swiss-roll µMFC achieves a unique combination of extreme miniaturization and high performance: with a footprint of only 4.16 $mm^2$ and a total active volume of ~800 nL, it is, to our knowledge, the smallest membrane-based MFC reported to date, occupying less than one-sixth the area of the closest competitor. Despite this ultra-compact geometry, it delivers competitive volumetric power densities of ~2.1 mW $cm^{-3}$ (total volume) and ~3.1 mW $cm^{-3}$ (anode volume), positioning its performance near the upper bound of the field.

This performance stems from the 3D architecture, which decouples device footprint from electrode area. Reducing spacer size or increasing windings presents a direct route to push volumetric density further, a scaling strategy unavailable to planar designs. A detailed comparison of the architectural approaches for these benchmarked devices is provided in Supplementary Table S2, highlighting the unique integration of a tunable membrane and dual-channel fluidics in our coaxial design.

Collectively, these results validate the coaxial Swiss-roll as a foundational architecture for integrated bio-electronics. This bottom-up, self-assembly platform provides a scalable microelectronic fabrication path to creating 3D, fluidically complex systems that are otherwise impractical to build, delivering a unique combination of ultra-small footprint, low volume, and high volumetric power density. By unifying a tunable ion-exchange membrane, high-surface-area electrodes, and a stability-enhancing operational scheme within a single monolithically integrated platform, this work moves beyond a standalone device.

The platform establishes a versatile and integrable blueprint not only for microbial fuel cells but for miniaturizing a broader class of electrochemical power sources. The same strain-engineered self-assembly and membrane-tuning approach could be adapted for direct methanol or ethanol fuel cells, where precise ionic management in three dimensions is equally critical.

Building on this performance foundation, we next examined the long-term stability and degradation mechanisms governing device lifetime.

**Fig. 6** Dual-mode operation and performance benchmarking of the coaxial Swiss-roll µMFC. **a** Concept of dual-mode operation. In Mode 1 (microbial), cells in an external reactor reduce the mediator, which transfers electrons to the coaxial anode while protons cross the PEM to the cathode. In Mode 2 (cell-free), cells are removed and the purified reduced mediator ($EM_{red}$) sustains anodic current; the oxidized mediator ($EM_{ox}$) can be recycled for regeneration. This strategy enables uninterrupted operation and prevents electrode fouling. **b** Experimental validation of dual-mode operation and PEM dependence. Polarization and power density curves show cell-free operation yields power densities comparable to microbial mode, with a peak output of ~1.7 μW and open-circuit voltages of 0.35–0.40 V (*i-ii*). PEM resistance strongly dictates performance: low-resistance PI membranes (~100 Ω) support currents up to ~8 μA and power maxima of ~2 μW, intermediate resistance (~500 Ω) reduces output to ~6 μA and ~1.3 μW, and high resistance (~10 kΩ) suppresses current to ~1 μA (*iii-iv*). Electrochemical impedance spectroscopy confirms stable, low impedance for membranes at 100–500 Ω and dominant capacitive behavior with large phase shifts at high resistance (~10 kΩ–1 MΩ), confirming transport-limited performance (*v-vi*). **c** Performance benchmarking against the state-of-the-art [16,32,33,47-50]. Ragone chart comparing volumetric power density against electrolyte volume for membrane-based μMFCs with total volumes <1 mL (Architectural details for each benchmarked device are summarized in Supplementary Table S2). Our coaxial Swiss-roll μMFC achieves ~2.1 mW $cm^{-3}$ (normalized to total volume, ~800 nL) and ~3.1 mW $cm^{-3}$ (normalized to anode volume, ~550 nL) within a footprint of 4.16 $mm^2$, ranking among the highest volumetric performers while maintaining the smallest footprint in the field.

### 3.7 Long-Term Stability and Degradation Mechanisms

The operational longevity of microscale bio-electrochemical systems is primarily limited by membrane fouling and interfacial degradation. In this work, organic fouling was defined as the accumulation and adsorption of organic molecules, here the soluble exogenous mediator (EM), onto the membrane surface and within near-surface proton-conducting pathways. In contrast, biofouling arises from microbial attachment, biofilm formation, and extracellular polymeric substance (EPS) deposition.

Fig. 7a schematically distinguishes three membrane states: clean electrolyte (i), microbial Mode 1 (biofouling + organic (EM) fouling) (ii), and cell-free Mode 2 (EM-only fouling) (iii). Corresponding SEM images and mediator crossover data are provided in Supplementary Fig. S13. SEM analysis reveals that Mode 1 results in extensive biomass coverage and surface obstruction, whereas Mode 2 preserves the intrinsic membrane topography, confirming the absence of microbial colonization under cell-free operation. Clean PI PEM exhibits characteristic dehydration-induced wrinkle morphology typical of free-standing thin films under compressive strain. After Mode 1, this intrinsic wrinkle topography disappears and the surface is fully masked by a continuous biofilm layer with visible cellular aggregates and debris.

The complete suppression of wrinkle features indicates uniform extracellular polymeric substance (EPS) coverage rather than structural smoothing of the PI membrane. In contrast, Mode 2 preserves the native wrinkle morphology with no detectable microbial structures, confirming that cell-free operation prevents biofilm formation inside the coaxial device.

Mediator crossover experiments (Supplementary Fig. S13b) demonstrate that optimally treated PI PEM (~270 Ω) effectively blocks EM transport over 24 h (below detection limit), while excessively hydrolyzed PI (~200 Ω) exhibits measurable crossover. This validates the selected 250–300 Ω resistance window as the optimal balance between proton conductivity and structural integrity.

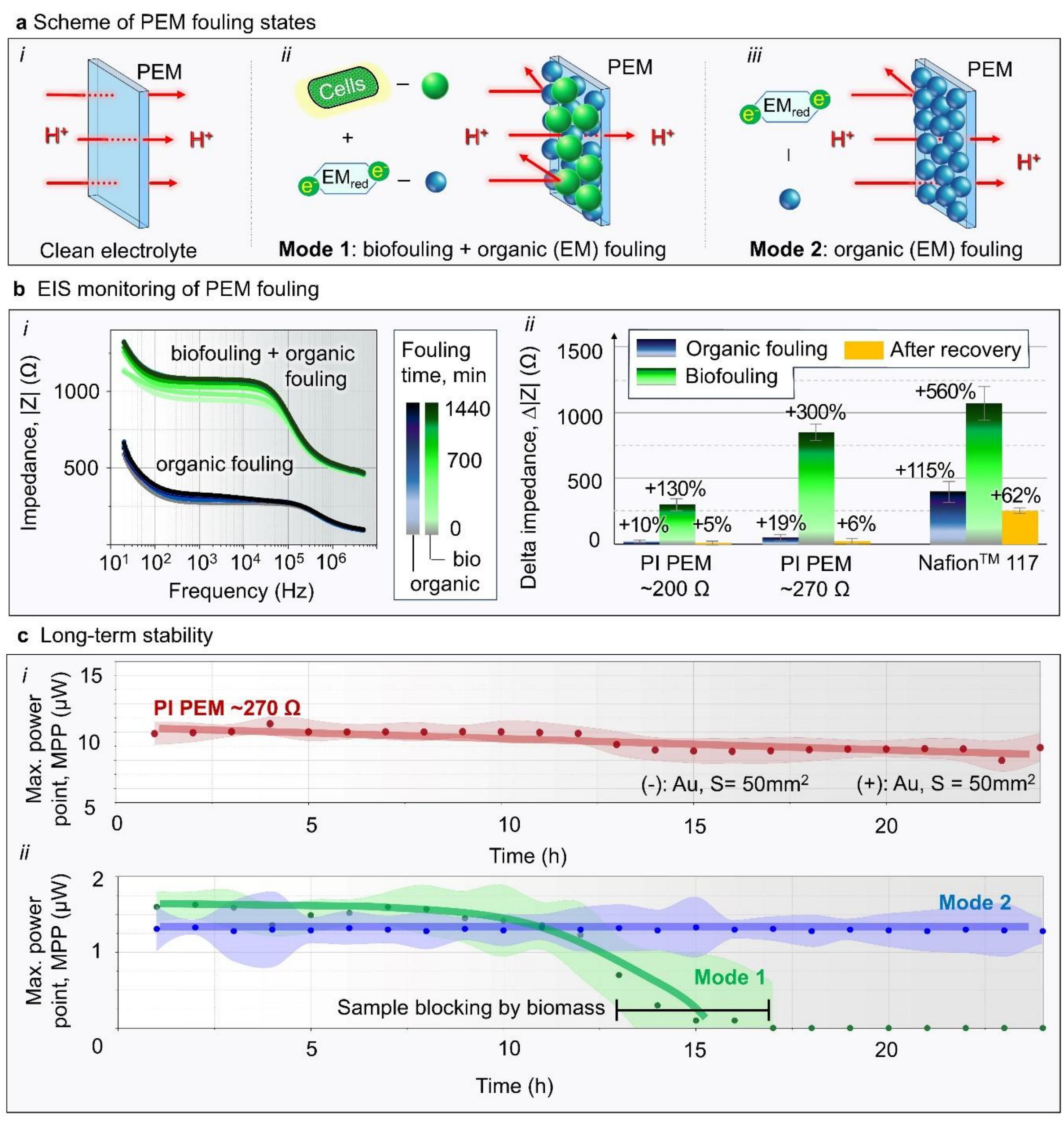

**Fig. 7** Mechanistic analysis of PEM fouling and long-term stability in the IFT-μMFC platform.

**a** Schematic illustration of membrane states: clean electrolyte (*i*), Mode 1 – combined biofouling and organic (EM) fouling (*ii*), and Mode 2 – organic fouling only under cell-free operation. $H^+$ transport occurs across the PEM, while mediator crossover and biomass accumulation depend on operational mode (*iii*). **b** Frequency-dependent impedance evolution of PI PEM during organic fouling and combined biofouling (*i*). Relative resistance increase ($\Delta|Z|$) after 24 h exposure, calculated at $10^3$ Hz for PI membranes and 500 Hz for Nafion™ 117 (membrane-dominated plateau region). Percentages are approximate and represent consistent trends (*ii*). Organic fouling alone induced modest increases for PI membranes (+10% and +19%) but a higher increase for Nafion™ 117 (+115%). Under biofouling, impedance rose significantly to +130% and +300% for PI membranes and +560% for Nafion™ 117. After flow-assisted recovery, PI membranes nearly returned to their initial resistance (+5–6% residual), whereas Nafion™ retained +62% elevated resistance, indicating irreversible fouling. **c** Long-term power stability. Component-level characterization of PI PEM showing stable maximum power over 24 h, with a minor decrease from ~10 μW to ~9 μW (i), consistent with biofouling (Fig. 7b). Integrated Swiss-roll IFT-μMFC under Mode 1 and Mode 2 operation (*ii*). Mode 2 maintains stable output (~1.2–1.4 μW), while Mode 1 exhibits progressive decline and failure after ~14–16 h due to biomass-induced channel blockage. The shaded area represents the error bars.

Electrochemical impedance spectroscopy further quantifies fouling (Fig. 7b and Supplementary Fig. S14). Organic fouling alone induces minor resistance increases for PI membranes (+10% for ~200 Ω; +19% for ~270 Ω). The relative resistance increase ($\Delta|Z|$) was calculated at $10^3$ Hz for PI membranes and 500 Hz for Nafion™ 117, frequencies corresponding to the membrane-dominated plateau region. Percentages are approximate and represent consistent trends across replicate measurements. The polyimide backbone, containing aromatic rings and imide groups, presents potential binding sites for such interactions through electrostatic and π-π stacking mechanisms [51]. Nafion™ 117 exhibited a substantial rise (+115%), consistent with its sulfonated, highly polar structure known to adsorb cationic dyes via electrostatic interactions and ion–dipole association [52,53]. Under biofouling, impedance increases dramatically: +130% (PI ~200 Ω), +300% (PI ~270 Ω), and +560% (Nafion™ 117). Biofouling of PEM represents a well-established failure mechanism in microbial fuel cells, where bacterial colonization and extracellular polymeric substance (EPS) accumulation impede proton transport and increase internal resistance [54,55]. Notably, after flow-assisted recovery, PI membranes nearly return to baseline (+5–6% residual), whereas Nafion™ retains +62% elevated resistance, indicating irreversible fouling due to strong dye–sulfonate binding and internal trapping within its hydrated ionic domains. These results highlight the superior recoverability and fouling tolerance of hydrolyzed PI compared to sulfonated fluoropolymers.

Long-term stability tests (Fig. 7c) confirm these findings. The standalone PI PEM module maintains stable peak power over 24 h, with a minor decrease from approximately 10 μW to 9 μW (Fig. 7c(i)). This slight decline is consistent with the gradual impedance increase observed under biofouling conditions (Fig. 7b), reflecting mild mediator and biomass adsorption on the membrane surface without compromising structural integrity. In the integrated Swiss-roll device, Mode 2 operation delivers stable output (~1.2–1.4 μW) with no degradation, whereas Mode 1 fails after approximately 14–16 h due to biomass-induced channel blockage (Fig. 7c(ii)). Importantly, microbes remain physically excluded from the electrochemical core and microelectronic interface in Mode 2, preventing biofilm-driven mechanical stress and catastrophic obstruction.

Collectively, these results demonstrate that biofouling, not mediator adsorption or intrinsic PI degradation, is the dominant failure mechanism in conventional integrated MFCs. The IFT-μMFC architecture eliminates this limitation through physical decoupling of microbial cultivation from electrochemical conversion. By confining biological activity to an external regeneration module, the platform provides a viable strategy to integrate microbial power generation with microscale electrochemical devices and future biohybrid microelectronics, leaving only mild, reversible organic fouling as the remaining limitation.

The identification of biofouling as the dominant failure mechanism, together with the excellent recoverability of optimized PI membranes, establishes clear design principles for future IFT-μMFC generations. Further enhancements could include surface charge engineering to electrostatically repel cationic mediators, zwitterionic coatings to resist microbial adhesion [56], and exploration of alternative mediator chemistries with reduced membrane affinity. The modular self-assembled coaxial architecture demonstrated here provides a promising platform for systematically implementing such advanced functionalities.

## 4. Conclusion

We have demonstrated a self-assembled coaxial Swiss-roll architecture that establishes a scalable platform for integrated bio-electronic microsystems. This approach synergistically combines a bottom-up, strain-engineered 3D microfabrication process with lithographically patterned and chemically tunable polyimide proton-exchange membranes. A dual-mode operational scheme decouples microbial metabolism from on-chip power generation, enhancing stability and mitigating biofouling. Long-term stability analysis reveals that biofouling, not chemical fouling or intrinsic membrane degradation, is the dominant failure mechanism in conventional architectures, while optimally tuned polyimide membranes combine high proton conductivity with effective mediator blocking and exhibit excellent recoverability after fouling.

The integrated platform achieves a high degree of miniaturization, operating with a total active volume of 0.80 µL and a compact footprint of 4.16 $mm^2$. It delivers volumetric power densities of ~3.1 mW $cm^{-3}$, positioning it at the forefront of miniaturized bio-electrochemical power. The programmability of the membrane and the inherent scalability of the architecture provide a direct route for further performance gains through geometric tuning.

This work provides a foundational bio-electronic platform. The co-fabrication of tunable ion-conducting membranes and high-surface-area 3D microelectrodes via self-assembly opens immediate application pathways in autonomous micro-robotics and drones, where such integrable, high-energy-density micro-power sources could significantly extend operational range. For environmental monitoring, the tubular architecture naturally lends itself to capillary-driven operation in passive soil or water sensors. Looking toward next-generation electronics, the ability to monolithically integrate energy harvesting with microfluidic channels suggests a compelling synergy with emerging chip-level liquid cooling systems for high-performance computing and AI data centers, potentially enabling simultaneous thermal management and waste-heat recovery within the same infrastructure. Future iterations could explore biodegradable polymer variants for environmentally benign applications. Ultimately, this work paves the way toward self-sustaining, intelligent microsystems that seamlessly bridge biological, electronic, and mechanical domains.

## Acknowledgements

Authors thank A. M. Placht, C. Schmidt, A. Dumler for the technical support. Authors thank Jie Lai (School of Materials Science and Engineering, Shenzhen University) for FTIR measurements. LM is a research productivity fellow supported by the Brazilian Council for Scientific and Technological Development (CNPq 312243/2025-1). D.K. and O.G.S. acknowledge support by the German Research Foundation DFG (KA 5051/3-1 SCHM 1298/32-1). DK acknowledges Horizon Europe HORIZON-EIC-2024-PATHFINDEROPEN-01, 101186701 — LEAF. O.G.S. acknowledges DFG Leibniz Programme.

## Authors' Contributions:

A.E., D.K. and O.G.S. conceived the idea and supervised the whole project. A.E., H.T. and D.D.K designed the experiment and fabricated the device. H.T., P.S., P.F., Y.L., L.M. assisted in the sample preparation, fabrication of experimental setup and testing the performance of the device. M.Z., D.K. and H.T. assisted in characterization of cell performance. A.E. and D.K. wrote paper draft. All authors discussed the results and contributed to the manuscript preparation.

## Declarations

## Conflict of Interest

The authors declare no competing interests

### Figure and table captions

**Fig. 1** Concept and realization of the self-assembled coaxial Swiss-roll bio-electronic platform. **a** Schematic of the 3D self-assembled architecture integrated into a microfluidic system, showing the inner catholyte channel (inner tube) and the outer anolyte channel (outer tube) with integrated parallel-plate electrodes (PPEs). **b** Device assembly roadmap. Four key stages: (1) Thin-film deposition of the multilayer 2D stack. (2) Self-assembly via selective sacrificial layer dissolution and hydrogel swelling. (3) Microfluidic integration by alignment and sealing of the rolled tube within a PDMS chip. (4) Plug-and-play operation by connection to external fluidic lines and reservoirs for power generation. **c** Schematic illustration of the indirect flow-type operating principle, in which externally processed anolyte is continuously supplied under flow to the coaxial microfluidic device for electrochemical energy conversion. This flow-through architecture decouples the local electrochemical kinetics from the external regeneration or charging processes. The platform supports a dual-mode operational scheme for functional versatility: a microbial mode 1 (functioning as a pure microbial fuel cell with integrated biology) and a cell-free mode 2 (operating as an abiotic electrochemical generator using pre-charged mediator). This allows the system to dynamically alternate between regenerative bio-power and stable electrochemistry, providing a robust interface between biological and electronic subsystems.

**Fig. 2** Structural characterization of the self-assembled coaxial Swiss-roll microtubes. Optical microscope images of the fabricated 3D architecture after release from the substrate. The main panel shows an overview of the structure, highlighting the simultaneous formation of the inner and outer tubes. The **left** panel provides a magnified view of the SU-8 spacer bars that define the interwinding gaps and fluidic channels. The **right** panel shows a higher-magnification side view of the tube entrance, revealing the tightly packed and uniform windings that ensure mechanical stability.

**Fig. 3** Microfluidic integration and electrical characterization of the coaxial platform. **a** Photographs of the device: the left image shows an open configuration with selective sealing of the inner and outer tubes, illustrating the preserved interwinding channels; the right image shows the fully assembled PDMS microfluidic chip with fluidic interconnects, **b** Electrical and microfluidic connection scheme, detailing the independent supply of fluids to the inner (catholyte) and outer (anolyte) channels, **c** Electrochemical impedance spectroscopy (EIS) performance of the integrated electrodes across a range of KCl solutions ($10^{-4}$ to 1 M). The data for the 3D interdigitated electrodes (IDEs, *i*), the 3D IDE–parallel-plate electrode (PPE) pair (*ii*), and the PPEs (*iii*) demonstrate that PPEs exhibit the lowest impedance, while the 3D IDEs provide stable performance across ionic strengths.

**Fig. 4** Enhanced proton transport via chemically tuned polyimide membranes in the Swiss-roll architecture. **a** Schematic cross-section illustrating proton transport pathways across single and multiple polyimide (PI) membrane layers in the multi-winding geometry. **b** SEM resolving the winding morphology, showing regions with single and multiple polyimide layers (corresponding to the pathways in a) and confirming the mechanically stable, uniform interwinding spacers. **c** Proton transfer dynamics, showing pH variation over time for treated PI membranes of different layer counts (1-4 layers) compared to commercial benchmarks (Nafion™ 117, Kapton™) after $H^+$ addition. **d** Chemical tuning and performance. (*i*) Mechanism of NaOH-induced hydrolysis, where partial imide ring-opening creates hydrophilic proton-conducting domains. (*ii*) Membrane resistance decreases with treatment time, defining an optimal window at ~250 Ω. (*iii*) Benchmarking shows a single tuned PI layer approaches Nafion™ 117 performance, compared to multi-layer PI stacks and Kapton™.

**Fig. 5** Component-level optimization of the bio-electrochemical system. Influence of electrode geometry (*i-ii*), exogenous mediator (EM) concentration (*iii-iv*), and proton exchange membrane (PEM) resistance (*v-vi*) on system performance. *i-ii,* Increasing anode surface area enhanced current and power, with large anodes (>200 mm$^2$) supporting >150 μA and >20 μW, while cathode enlargement yielded minimal gains. *iii-iv*, Mediator concentration exhibited an optimal window (~0.0125–0.025 M); lower concentrations limited electron transfer while higher levels induced cytotoxicity. *v-vi*, PEM resistance was critical: low-resistance PI membranes (~200 Ω) enabled high performance, whereas intermediate (~500 Ω) and high ($\geqslant$ 9 kΩ) resistance severely limited output. These results establish key design rules: maximize anode area, maintain mediator within a biocompatible range, and minimize PEM resistance. Data points are measurements; lines are guides for the eye. (*ii-vi*).

**Fig. 6** Dual-mode operation and performance benchmarking of the coaxial Swiss-roll μMFC. **a** Concept of dual-mode operation. In Mode 1 (microbial), cells in an external reactor reduce the mediator, which transfers electrons to the coaxial anode while protons cross the PEM to the cathode. In Mode 2 (cell-free), cells are removed and the purified reduced mediator ($EM_{red}$) sustains anodic current; the oxidized mediator ($EM_{ox}$) can be recycled for regeneration. This strategy enables uninterrupted operation and prevents electrode fouling. **b** Experimental validation of dual-mode operation and PEM dependence. Polarization and power density curves show cell-free operation yields power densities comparable to microbial mode, with a peak output of ~1.7 μW and open-circuit voltages of 0.35–0.40 V (*i-ii*). PEM resistance strongly dictates performance: low-resistance PI membranes (~100 Ω) support currents up to ~8 μA and power maxima of ~2 μW, intermediate resistance (~500 Ω) reduces output to ~6 μA and ~1.3 μW, and high resistance (~10 kΩ) suppresses current to ~1 μA (*iii-iv*). Electrochemical impedance spectroscopy confirms stable, low impedance for membranes at 100–500 Ω and dominant capacitive behavior with large phase shifts at high resistance (~10 kΩ–1 MΩ), confirming transport-limited performance (*v-vi*). **c** Performance benchmarking against the state-of-the-art [16,32,33,47-50]. Ragone chart comparing volumetric power density against electrolyte volume for membrane-based μMFCs with total volumes <1 mL (Architectural details for each benchmarked device are summarized in Supplementary Table S2). Our coaxial Swiss-roll μMFC achieves ~2.1 mW cm$^{-3}$ (normalized to total volume, ~800 nL) and ~3.1 mW cm$^{-3}$ (normalized to anode volume, ~550 nL) within a footprint of 4.16 mm$^2$, ranking among the highest volumetric performers while maintaining the smallest footprint in the field.

**Fig. 7** Mechanistic analysis of PEM fouling and long-term stability in the IFT-μMFC platform. **a** Schematic illustration of membrane states: clean electrolyte (*i*), Mode 1 – combined biofouling and organic (EM) fouling (*ii*), and Mode 2 – organic fouling only under cell-free operation. $H^+$ transport occurs across the PEM, while mediator crossover and biomass accumulation depend on operational mode (*iii*). **b** Frequency-dependent impedance evolution of PI PEM during organic fouling and combined biofouling (*i*). Relative resistance increase ($\Delta|Z|$) after 24 h exposure, calculated at $10^3$ Hz for PI membranes and 500 Hz for Nafion™ 117 (membrane-dominated plateau region). Percentages are approximate and represent consistent trends (*ii*). Organic fouling alone induced modest increases for PI membranes (+10% and +19%) but a higher increase for Nafion™ 117 (+115%). Under biofouling, impedance rose significantly to +130% and +300% for PI membranes and +560% for Nafion™ 117. After flow-assisted recovery, PI membranes nearly returned to their initial resistance (+5–6% residual), whereas Nafion™ retained +62% elevated resistance, indicating irreversible fouling. **c** Long-term power stability. Component-level characterization of PI PEM showing stable maximum power over 24 h, with a minor decrease from ~10 μW to ~9 μW (i), consistent with biofouling (Fig. 7b). Integrated Swiss-roll IFT-μMFC under Mode 1 and Mode 2 operation (*ii*). Mode 2 maintains stable output (~1.2–1.4 μW), while Mode 1 exhibits progressive decline and failure after ~14–16 h due to biomass-induced channel blockage. The shaded area represents the error bars.

## References

[1] L. R. Gomez Palacios, A. G. Bracamonte, Development of nano- and microdevices for the next generation of biotechnology, wearables and miniaturized instrumentation. RSC Adv. 12(20), 12806–12822 (2022). https://doi.org/10.1039/D2RA02008D

[2] A. T. Kutbee, R. R. Bahabry, K. O. Alamoudi, M. T. Ghoneim, M. D. Cordero, et al., Flexible and biocompatible high-performance solid-state micro-battery for implantable orthodontic system. npj Flex. Electron. 1(1), 7 (2017). https://doi.org/10.1038/s41528-017-0008-7

[3] J. Ni, A. Dai, Y. Yuan, L. Li, J. Lu, Three-Dimensional Microbatteries beyond Lithium Ion. Matter 2(6), 1366–1376 (2020). https://doi.org/10.1016/j.matt.2020.04.020

[4] C. Santoro, F. Soavi, A. Serov, C. Arbizzani, P. Atanassov, Self-powered

supercapacitive microbial fuel cell: The ultimate way of boosting and harvesting power. Biosens. Bioelectron. 78 229–235 (2016). https://doi.org/10.1016/j.bios.2015.11.026

[5] Y. Tekle, A. Demeke, Review on microbial fuel cell. Int. J. Eng. Technol. 8, 424–427 (2015)

[6] J. Ma, J. Zhang, Y. Zhang, Q. Guo, T. Hu, et al., Progress on anodic modification materials and future development directions in microbial fuel cells. J. Power Sources 556 232486 (2023). https://doi.org/10.1016/j.jpowsour.2022.232486

[7] J. M. Moradian, Z. Fang, Y.-C. Yong, Recent advances on biomass-fueled microbial fuel cell. Bioresour. Bioprocess. 8(1), 14 (2021). https://doi.org/10.1186/s40643-021-00365-7

[8] Y. Cao, H. Mu, W. Liu, R. Zhang, J. Guo, et al., Electricigens in the anode of microbial fuel cells: pure cultures versus mixed communities. Microb. Cell Fact. 18(1), 39 (2019). https://doi.org/10.1186/s12934-019-1087-z

[9] P. Parkhey, R. Sahu, Microfluidic microbial fuel cells: Recent advancements and future prospects. Int. J. Hydrogen Energy 46(4), 3105–3123 (2021). https://doi.org/10.1016/j.ijhydene.2020.07.019

[10] B. E. Logan, M. J. Wallack, K. Y. Kim, W. He, Y. Feng, et al., Assessment of Microbial Fuel Cell Configurations and Power Densities. Environ. Sci. Technol. Lett. 2(8), 206–214 (2015). https://doi.org/10.1021/acs.estlett.5b00180

[11] O. A. Ibrahim, M. Navarro-Segarra, P. Sadeghi, N. Sabaté, J. P. Esquivel, et al., Microfluidics for Electrochemical Energy Conversion. Chem. Rev. 122(7), 7236–7266 (2022). https://doi.org/10.1021/acs.chemrev.1c00499

[12] L. Gong, M. Abbaszadeh Amirdehi, J. M. Sonawane, N. Jia, L. Torres de Oliveira, et al., Mainstreaming microfluidic microbial fuel cells: a biocompatible membrane grown in situ improves performance and versatility. Lab Chip 22(10), 1905–1916 (2022). https://doi.org/10.1039/D2LC00098A

[13] M. A. Amirdehi, N. Khodaparastasgarabad, H. Landari, M. P. Zarabadi, A. Miled, et al., A High-Performance Membraneless Microfluidic Microbial Fuel Cell for Stable, Long-Term Benchtop Operation Under Strong Flow. ChemElectroChem 7(10), 2227–2235 (2020). https://doi.org/10.1002/celc.202000040

[14] W. Yang, K. K. Lee, S. Choi, A laminar-flow based microbial fuel cell array. Sensors Actuators B Chem. 243 292–297 (2017). https://doi.org/10.1016/j.snb.2016.11.155

[15] K. Scott, Membranes and separators for microbial fuel cells. in *Microbial*

*Electrochemical and Fuel Cells* 153–178 (Elsevier, 2016). doi:10.1016/B978-1-78242-375-1.00005-8. https://doi.org/10.1016/B978-1-78242-375-1.00005-8

[16] F. Qian, M. Baum, Q. Gu, D. E. Morse, A 1.5 µL microbial fuel cell for on-chip bioelectricity generation. Lab Chip 9(21), 3076 (2009). https://doi.org/10.1039/b910586g

[17] C. A. Machado, G. O. Brown, R. Yang, T. E. Hopkins, J. G. Pribyl, et al., Redox Flow Battery Membranes: Improving Battery Performance by Leveraging Structure–Property Relationships. ACS Energy Lett. 6(1), 158–176 (2021). https://doi.org/10.1021/acsenergylett.0c02205

[18] D. Düerkop, H. Widdecke, C. Schilde, U. Kunz, A. Schmiemann, Polymer Membranes for All-Vanadium Redox Flow Batteries: A Review. Membranes (Basel). 11(3), 214 (2021). https://doi.org/10.3390/membranes11030214

[19] G. L. Soloveichik, Flow Batteries: Current Status and Trends. Chem. Rev. 115(20), 11533–11558 (2015). https://doi.org/10.1021/cr500720t

[20] K. B. Lam, E. F. Irwin, K. E. Healy, L. Lin, Bioelectrocatalytic self-assembled thylakoids for micro-power and sensing applications. Sensors Actuators B Chem. 117(2), 480–487 (2006). https://doi.org/10.1016/j.snb.2005.12.057

[21] K. B. Lam, E. A. Johnson, M. Chiao, L. Lin, A MEMS Photosynthetic Electrochemical Cell Powered by Subcellular Plant Photosystems. J. Microelectromechanical Syst. 15(5), 1243–1250 (2006). https://doi.org/10.1109/JMEMS.2006.880296

[22] M. Chiao, K. B. Lam, L. Lin, Micromachined microbial and photosynthetic fuel cells. J. Micromechanics Microengineering 16(12), 2547–2553 (2006). https://doi.org/10.1088/0960-1317/16/12/005

[23] D. Karnaushenko, T. Kang, V. K. Bandari, F. Zhu, O. G. Schmidt, 3D Self-Assembled Microelectronic Devices: Concepts, Materials, Applications. Adv. Mater. 1902994 1–30 (2019). https://doi.org/10.1002/adma.201902994

[24] S. Liu, L. Wang, B. Zhang, B. Liu, J. Wang, et al., Novel sulfonated polyimide/polyvinyl alcohol blend membranes for vanadium redox flow battery applications. J. Mater. Chem. A 3(5), 2072–2081 (2015). https://doi.org/10.1039/C4TA05504G

[25] X. Huang, S. Zhang, Y. Zhang, H. Zhang, X. Yang, Sulfonated polyimide/chitosan composite membranes for a vanadium redox flow battery: influence of the sulfonation degree of the sulfonated polyimide. Polym. J. 48(8), 905–918 (2016). https://doi.org/10.1038/pj.2016.42

[26] J. Li, X. Yuan, S. Liu, Z. He, Z. Zhou, et al., A Low-Cost and High-Performance Sulfonated Polyimide Proton-Conductive Membrane for Vanadium Redox Flow/Static Batteries. ACS Appl. Mater. Interfaces 9(38), 32643–32651 (2017). https://doi.org/10.1021/acsami.7b07437

[27] W. Xu, J. Long, J. Liu, H. Luo, H. Duan, et al., A novel porous polyimide membrane with ultrahigh chemical stability for application in vanadium redox flow battery. Chem. Eng. J. 428 131203 (2022). https://doi.org/10.1016/j.cej.2021.131203

[28] D. Karnaushenko, N. Münzenrieder, D. D. Karnaushenko, B. Koch, A. K. Meyer, et al., Biomimetic Microelectronics for Regenerative Neuronal Cuff Implants. Adv. Mater. 27(43), 6797–6805 (2015). https://doi.org/10.1002/adma.201503696

[29] D. D. Karnaushenko, D. Karnaushenko, D. Makarov, O. G. Schmidt, Compact helical antenna for smart implant applications. NPG Asia Mater. 7(6), e188–e188 (2015). https://doi.org/10.1038/am.2015.53

[30] A. I. Egunov, Z. Dou, D. D. Karnaushenko, F. Hebenstreit, N. Kretschmann, et al., Impedimetric Microfluidic Sensor-in-a-Tube for Label-Free Immune Cell Analysis. Small 17(5), (2021). https://doi.org/10.1002/smll.202002549

[31] E. Ghosh, A. I. Egunov, D. Karnaushenko, M. Medina-Sánchez, O. G. Schmidt, Self-assembled sensor-in-a-tube as a versatile tool for label-free EIS viability investigation of cervical cancer cells. Frequenz 76(11–12), 729–740 (2022). https://doi.org/10.1515/freq-2022-0090

[32] S. Choi, H.-S. Lee, Y. Yang, P. Parameswaran, C. I. Torres, et al., A μL-scale micromachined microbial fuel cell having high power density. Lab Chip 11(6), 1110 (2011). https://doi.org/10.1039/c0lc00494d

[33] D. Vigolo, T. T. Al-Housseiny, Y. Shen, F. O. Akinlawon, S. T. Al-Housseiny, et al., Flow dependent performance of microfluidic microbial fuel cells. Phys. Chem. Chem. Phys. 16(24), 12535 (2014). https://doi.org/10.1039/c4cp01086h

[34] O. G. Schmidt, K. Eberl, Thin solid films roll up into nanotubes. Nature 410(6825), 168–168 (2001). https://doi.org/10.1038/35065525

[35] V. Y. Prinz, V. Seleznev, A. . Gutakovsky, A. . Chehovskiy, V. Preobrazhenskii, et al., Free-standing and overgrown InGaAs/GaAs nanotubes, nanohelices and their arrays. Phys. E Low-dimensional Syst. Nanostructures 6(1–4), 828–831 (2000). https://doi.org/10.1016/S1386-9477(99)00249-0

[36] J. Wang, D. Karnaushenko, M. Medina-Sánchez, Y. Yin, L. Ma, et al., Three-Dimensional Microtubular Devices for Lab-on-a-Chip Sensing Applications. ACS

Sensors 4(6), 1476–1496 (2019). https://doi.org/10.1021/acssensors.9b00681
[37] P. Lepucki, A. I. Egunov, M. Rosenkranz, R. Huber, A. Mirhajivarzaneh, et al., Self-Assembled Rolled-Up Microcoils for nL Microfluidics NMR Spectroscopy. Adv. Mater. Technol. 6(1), 2000679 (2021). https://doi.org/10.1002/admt.202000679
[38] Y. Lee, V. K. Bandari, Z. Li, M. Medina-Sánchez, M. F. Maitz, et al., Nano-biosupercapacitors enable autarkic sensor operation in blood. Nat. Commun. 12(1), 4–13 (2021). https://doi.org/10.1038/s41467-021-24863-6
[39] F. Gabler, D. D. Karnaushenko, D. Karnaushenko, O. G. Schmidt, Magnetic origami creates high performance micro devices. Nat. Commun. 10(1), 3013 (2019). https://doi.org/10.1038/s41467-019-10947-x
[40] C. N. Saggau, F. Gabler, D. D. Karnaushenko, D. Karnaushenko, L. Ma, et al., Wafer-Scale High-Quality Microtubular Devices Fabricated via Dry-Etching for Optical and Microelectronic Applications. Adv. Mater. 32(37), 2003252 (2020). https://doi.org/10.1002/adma.202003252
[41] F. Li, J. Wang, L. Liu, J. Qu, Y. Li, et al., Self-Assembled Flexible and Integratable 3D Microtubular Asymmetric Supercapacitors. Adv. Sci. 6(20), 1901051 (2019). https://doi.org/10.1002/advs.201901051
[42] Z. Qu, M. Zhu, Y. Yin, Y. Huang, H. Tang, et al., A Sub-Square-Millimeter Microbattery with Milliampere-Hour-Level Footprint Capacity. Adv. Energy Mater. 12(28), 2200714 (2022). https://doi.org/10.1002/aenm.202200714
[43] H. Tang, L. M. M. Ferro, D. D. Karnaushenko, O. Selyshchev, X. Wang, et al., An “Ion Harvester” Battery in Soil Empowered by a Microfluidic Pump and Interlayer Confinement within Micro-Swiss-Rolls. Adv. Funct. Mater. 35(14), (2025). https://doi.org/10.1002/adfm.202418872
[44] M. Zhu, J. Hu, Q. Lu, H. Dong, D. D. Karnaushenko, et al., A Patternable and In Situ Formed Polymeric Zinc Blanket for a Reversible Zinc Anode in a Skin-Mountable Microbattery. Adv. Mater. 33(8), 2007497 (2021). https://doi.org/10.1002/adma.202007497
[45] D. Karnaushenko, T. Kang, O. G. Schmidt, Shapeable Material Technologies for 3D Self-Assembly of Mesoscale Electronics. Adv. Mater. Technol. 4(4), 1–29 (2019). https://doi.org/10.1002/admt.201800692
[46] X. Lai, M. Yang, H. Wu, D. Li, Modular Microfluidics: Current Status and Future Prospects. Micromachines 13(8), 1–22 (2022). https://doi.org/10.3390/mi13081363
[47] H. Ren, C. I. Torres, P. Parameswaran, B. E. Rittmann, J. Chae, Improved current and

power density with a micro-scale microbial fuel cell due to a small characteristic length. Biosens. Bioelectron. 61 587–592 (2014). https://doi.org/10.1016/j.bios.2014.05.037

[48] H. Jiang, M. A. Ali, Z. Xu, L. J. Halverson, L. Dong, Integrated Microfluidic Flow-Through Microbial Fuel Cells. Sci. Rep. 7(1), 41208 (2017). https://doi.org/10.1038/srep41208

[49] J. E. Mink, J. P. Rojas, B. E. Logan, M. M. Hussain, Vertically Grown Multiwalled Carbon Nanotube Anode and Nickel Silicide Integrated High Performance Microsized (1.25 μL) Microbial Fuel Cell. Nano Lett. 12(2), 791–795 (2012). https://doi.org/10.1021/nl203801h

[50] B. Şen-Doğan, M. Okan, N. Afşar-Erkal, E. Özgür, Ö. Zorlu, et al., Enhancement of the Start-Up Time for Microliter-Scale Microbial Fuel Cells (μMFCs) via the Surface Modification of Gold Electrodes. Micromachines 11(7), 703 (2020). https://doi.org/10.3390/mi11070703

[51] A. Murugesan, P. Mahendran, High-Performance Polyimides with Pendant Fluorenylidene Groups: Synthesis, Characterization and Adsorption Behaviour. J. Polym. Environ. 28(9), 2393–2409 (2020). https://doi.org/10.1007/s10924-020-01777-w

[52] K. A. Mauritz, R. B. Moore, State of Understanding of Nafion. Chem. Rev. 104(10), 4535–4586 (2004). https://doi.org/10.1021/cr0207123

[53] A. Kusoglu, A. Z. Weber, New Insights into Perfluorinated Sulfonic-Acid Ionomers. Chem. Rev. 117(3), 987–1104 (2017). https://doi.org/10.1021/acs.chemrev.6b00159

[54] S. G. A. Flimban, S. H. A. Hassan, M. M. Rahman, S.-E. Oh, The effect of Nafion membrane fouling on the power generation of a microbial fuel cell. Int. J. Hydrogen Energy 45(25), 13643–13651 (2020). https://doi.org/10.1016/j.ijhydene.2018.02.097

[55] L. Zang, X.-L. Yang, H. Xu, Y.-G. Xia, H.-L. Song, A novel integrated microbial fuel cell-membrane bioreactor (MFC-MBR) for controlling the spread of antibiotic and antibiotic resistance genes while simultaneously alleviating membrane fouling. Chem. Eng. J. 487 150578 (2024). https://doi.org/10.1016/j.cej.2024.150578

[56] G. Gu, X. Yang, Y. Li, J. Guo, J. Huang, et al., Advanced zwitterionic polymeric membranes for diverse applications beyond antifouling. Sep. Purif. Technol. 356 129848 (2025). https://doi.org/10.1016/j.seppur.2024.129848

## Supporting materials

# Thin-Film-Engineered Self-Assembly of 3D Coaxial Microfluidics with a Tunable Polyimide Membrane for Bioelectronic Power

Aleksandr I. Egunov[1]*, Hongmei Tang[1], Pablo E. Saenz[1], Dmitriy D. Karnaushenko[1], Yumin Luo[1,2], Chao Zhong[3], Xinyu Wang[3], Yang Huang[4], Pavel Fedorov[1], Leandro Merces[1,5], Minshen Zhu[1]*, Daniil Karnaushenko[1]* and Oliver, G. Schmidt[1,6,7]*

[1] Research Center for Materials, Architectures and Integration of Nanomembranes, Chemnitz University of Technology, Chemnitz, Germany.

[2] Chair of Microsystems Technology, Ruhr University Bochum, Bochum, Germany

[3] Shenzhen Key Laboratory of Materials Synthetic Biology, Key Laboratory of Quantitative Synthetic Biology, Shenzhen Institute of Synthetic Biology, Shenzhen Institutes of Advanced Technology, Chinese Academy of Sciences, Shenzhen, China.

[4] Departament of Physics and Materials Science, City Univercity of Hong Kong, Hong Kong, China.

[5] Ilum School of Science, Brazilian Center for Research in Energy and Materials, Campinas, Brazil.

[6] Material Systems for Nanoelectronics, Chemnitz University of Technology, Chemnitz, Germany

[7] Nanophysics, Dresden University of Technology, Dresden, Germany

*Corresponding authors

E-mail: aleksandr.egunov@main.tu-chemnitz.de,

E-mail: minshen.zhu@main.tu-chemnitz.de,

E-mail: daniil.karnaushenko@main.tu-chemnitz.de,

E-mail: oliver.schmidt@main.tu-chemnitz.de.

**Supplementary information**

Extended data for Fig. 3, Fig. S6 and Fig. S7

The observed impedance and capacitance behavior (Fig. S6) can be directly explained by the physics of electric double-layer (EDL) formation. At low ionic strength ($10^{-5}$–$10^{-3}$ M KCl), the Debye screening length ($\lambda_D$) (Supplementary Table S1) is extended, reaching ~100 nm at $10^{-5}$ M and ~10 nm at $10^{-3}$ M. In this regime, counterion accumulation at the electrode is incomplete, resulting in low double-layer capacitance and high impedance. This effect is particularly pronounced in 2D IDEs, where limited effective surface area and larger inter-electrode distances restrict efficient charge compensation.

At intermediate concentrations ($10^{-2}$–$10^{-1}$ M), $\lambda_D$ contracts to ~3–1 nm, enabling more effective EDL charging and a marked increase in capacitance. In this range, the coaxial geometry of 3D IDEs provides a decisive advantage: the increased interfacial area and shortened ionic pathways reduce impedance significantly compared to 2D IDEs. PPEs, by contrast, offer extended planar interfaces that sustain strong and uniform EDL formation, leading to the lowest observed impedance across the frequency range.

At high ionic strength (1 M KCl), $\lambda_D$ falls below 1 nm, producing saturated double-layer capacitance at all electrode types. Here, impedance is dominated by bulk electrolyte conductivity, and the performance gap between electrode architectures diminishes. Nonetheless, PPEs maintain the most stable capacitive response, while 3D IDEs preserve broader frequency-dependent charge storage capacity.

These findings confirm that coaxial self-rolled electrodes provide favorable conditions for EDL formation across a broad concentration range, combining low impedance with stable capacitive behavior. Such characteristics are critical for microbial fuel cell operation, where efficient charge transfer and minimized resistive losses directly translate into improved bioelectrochemical performance.

Ionic strength dependence (KCl solutions, Fig. 3c and Fig. S6).

At 1 M KCl, PPEs exhibited impedance values as low as ~$1.5 \times 10^3$ Ω at 100 Hz, compared to ~$1\times10^5$ Ω for 2D IDEs and ~$1 \times 10^4$ Ω for 3D IDEs. At lower concentration ($10^{-5}$ M KCl), impedance increased to ~$5\times10^6$ Ω for 2D IDEs, ~$1 \times 10^6$ Ω for 3D IDEs, and ~$3\times10^5$ Ω for PPEs.

This reduction can be described by the classical Debye screening relation:

$$\boldsymbol{\lambda_D = \sqrt{\frac{\varepsilon k_B T}{2 N_A e^2 I}}}$$

where $\lambda_D$ is the Debye length, $\varepsilon$ is the dielectric permittivity of water ($\varepsilon \approx 80$),

$k_B$ is Boltzmann constant = $1.380\times10^{-23}$ J·K$^{-1}$ , $T$ — Absolute temperature, assumed 298 K (25 °C), $N_A$ — Avogadro's number = $6.022\times10^{23}$ mol$^{-1}$, $I$ — Ionic strength (mol·L$^{-1}$), defined as

$$\boldsymbol{I = \frac{1}{2}\sum C_i Z_i^2}$$

where $C_i$ is molar concentration of ion i, and $Z_i$ its valence.

Increasing $I$ from $10^{-5}$ M to 1 M reduces $\lambda_D$ from ~100 nm to ~0.3 nm, explaining the improved charge screening and lowered impedance. Capacitance estimates from the Z′–C′ plots confirm this: PPEs reach ~$3\times10^{-9}$ F at 1 M KCl, compared to ~$2\times10^{-10}$ F for 2D IDEs under the same conditions.

Dielectric dependence (ethanol–water mixtures, Fig. S7).

At fixed ionic content, decreasing the solvent dielectric constant from $\varepsilon \approx 80$ (pure water) to $\varepsilon \approx 25$ (pure ethanol) systematically increased impedance. For PPEs, |Z| at 100 Hz rose from ~$1\times10^4$ Ω (water) to ~$1\times10^7$ Ω (ethanol), i.e. a three-order-of-magnitude shift.

The capacitance followed the expected scaling:

$$\boldsymbol{C \sim \frac{\varepsilon}{\lambda_D}}$$

where $\lambda_D$ remains constant at fixed ionic concentration. This results in a direct proportionality of capacitance with ε. Experimentally, PPE capacitance decreased from ~$2\times10^{-9}$ F (water) to ~$7\times10^{-11}$ F (ethanol).

2D IDEs displayed smaller changes (~$1\times10^{-10}$ F to ~$5\times10^{-11}$ F) because of their strongly localized field lines, while 3D IDEs showed intermediate sensitivity.

Extended data for Fig. S8

## Proton exchange in alkali-treated polyimide membranes

The alkali (NaOH) treatment of PI PEM partially hydrolyzes the imide rings of polyimide (PI) into their open-chain poly(amic acid) form. This transformation introduces carboxylate (–$COO^{-}$) and amide (–CONH–) groups that enhance ionic transport and water affinity, beneficial for proton conduction. The reaction proceeds via nucleophilic attack of hydroxide ions on the imide carbonyl carbon, followed by ring-opening to yield amic acid intermediates.

In the FTIR spectra (Fig. S9a), pristine PI shows characteristic imide peaks at:
1775 $cm^{-1}$ (C=O asymmetric stretch)
1720 $cm^{-1}$ (C=O symmetric stretch)
1380 $cm^{-1}$ (C–N stretch)
720 $cm^{-1}$ (imide ring deformation)

After NaOH exposure, new and shifted features appear:

Broad O–H/N–H stretching (3200–3600 $cm^{-1}$): from hydroxyl and amide groups in poly(amic acid).
Weakened imide C=O peaks (1775 and 1720 $cm^{-1}$): indicating ring cleavage.
New amide C=O band (~1650 $cm^{-1}$): characteristic of amic acid formation.
New $COO^{-}$ symmetric stretching (~1420 $cm^{-1}$): confirming partial conversion to carboxylate.

The improvement in ionic transport after alkali treatment is consistent with hydrolysis of imide groups in PI:

$$-\mathrm{CO{-}NH{-}CO}- \;\rightarrow\; -\mathrm{COOH} + -\mathrm{CONH}- \text{ or } -\mathrm{COO^{-}NA^{+}} \,.$$

The repeat unit of the PI contains two imide linkages per monomer, which are the primary sites for hydroxide attack. Controlled NaOH hydrolysis converts a fraction of imide rings into poly(amic)-like structures bearing amide and carboxylate/carboxylic groups. In FTIR this is manifested by a reduced and broadened imide C=O envelope (~1775 and 1720 $cm^{-1}$), a shoulder/increase around ~1650 $cm^{-1}$ (amide C=O / amic acid), increased O–H/N–H absorption (~3400 $cm^{-1}$), and modest growth near 1420 $cm^{-1}$ ($COO^{-}$ symmetric). Excessive hydrolysis produces a high density of hydrophilic groups, causing strong swelling, gelation and partial dissolution in aqueous media, a condition we avoid by selecting an intermediate target resistance (~250 Ω at $10^{3}$ Hz).

Electrochemical impedance spectroscopy (EIS) was used to quantify membrane resistance as a function of alkali treatment time. The total impedance is represented as

$$Z(\omega) \approx Z'(\omega) + jZ''(\omega)$$

where $Z'(\omega)$ is the real component (resistive contribution) and $Z''(\omega)$ is the imaginary component (capacitive contribution). For thin membranes, the dominant parameter is the real resistance at mid frequencies ($10^2$–$10^4$ Hz), where electrode polarization is negligible and bulk ionic conduction dominates.

The membrane resistance was extracted from the Bode plot at $10^3$ Hz,

$$R_{mem} \approx Z'(\omega = 10^3 Hz)$$

giving values of ~$10^6$ Ω for pristine PI, ~$10^5$ Ω after $10^2$ min of NaOH treatment, and ~$10^3$ Ω after $10^3$ min (Fig. S9b). At very long treatment times (>7.5 × $10^3$ min), $R_{mem}$ increased again due to membrane fragilisation and partial loss of structural integrity.

While Nafion™ 117 exhibited the lowest resistance (~$10^2$ Ω), treated PI reached comparable performance at optimal treatment times ($10^3$–5x$10^3$ min), with a significant mechanical advantage in multilayer coaxial integration. In contrast, Kapton™ remained insulating with $R_{mem}$>$10^7$ Ω

Thus, alkali-treated PI provides a tunable balance between ionic conductivity and structural robustness, making it particularly suited for Swiss-roll µMFC architectures where membrane winding introduces regions of reduced effective thickness.

**Supplementary Table 1. Calculated Debye screening length ($\lambda_D$) for KCl solutions**

| KCl Concentration (M) | Debye Length, $\lambda_D$ (nm) |
|---|---|
| $10^{-5}$ M | 96.3 |
| $10^{-4}$ M | 30.5 |
| $10^{-3}$ M | 9.63 |
| $10^{-2}$ M | 3.05 |
| $10^{-1}$ M | 0.96 |
| 1 M | 0.3 |

**Supplementary Table 2. Architectural and functional comparison of membrane-based sub-1 mL µMFCs**

| Ref. | Anolyte volume (A.v.), nL. | Total volume (T.v.), nL. | Architecture/Fabrication | Membrane/Separator | Maximum volumetric power density normalized per A.v., $mW/cm^3$ | Maximum volumetric power density normalized per T.v., $mW/cm^3$ | Key features/ Functional notes |
|---|---|---|---|---|---|---|---|
| Ref [16] | 1500 | 5500 | Vertically stacked dual-chamber, microfluidic | Nafion™ 117 | 0.015 | 0.004 | Early on-chip µMFC demonstration. |
| Ref [47] | 100000 | 200000 | Planar, glass slide assembly, gasket-defined | Nafion™ 117 | 3.3 | 1.65 | Focus on characteristic length; planar chambers. |
| Ref [32] | 4500 | 9000 | Planar MEMS, PDMS-defined chambers | CMI 7000 | 2.3 | 1.15 | High power density; uses conductive Geobacter biofilm; high coulombic efficiency. |
| [33] | 5000 | n.a. | Planar, two-compartment flow cell, | Nafion™ 117 | 0.9 | n.a | Continuous flow operation; serpentine channel design; carbon paper electrodes. |
| | | | serpentine channels | | | | |
| Ref [48] | 58800 | n.a. | Microfluidic flow-through with 3D graphene foam anode | Nafion™ 117 | 0.745 | n.a. | Flow-through anode chamber; 3D graphene foam for intimate microbe interaction; fast mass transport. |
| [49] | 1250 | n.a. | Planar, Si-based with vertical CNT anode | Nafion™ 117 | 0.392 | n.a. | High surface-area vertically aligned carbon nanotubes anode; compact total anolyte volume (1.25 μL). |
| Ref [50] | 10400 | 20800 | Planar MEMS, silicon double-chamber | Nafion™ 117 | 0.315 and 0.33 | 0.156 and 0.165 | Focus on anode surface modification (thiol SAMs) to reduce start-up time. |
| This work | 550 | 800 | Coaxial Swiss-roll, strain-driven self-assembly | Tunable Polyimide | 3.1 | 2.1 | Coaxial channels; integrated & tunable PEM; dual-mode operation; monolithic 3D integration. |

## Supplementary Figs. 1–11

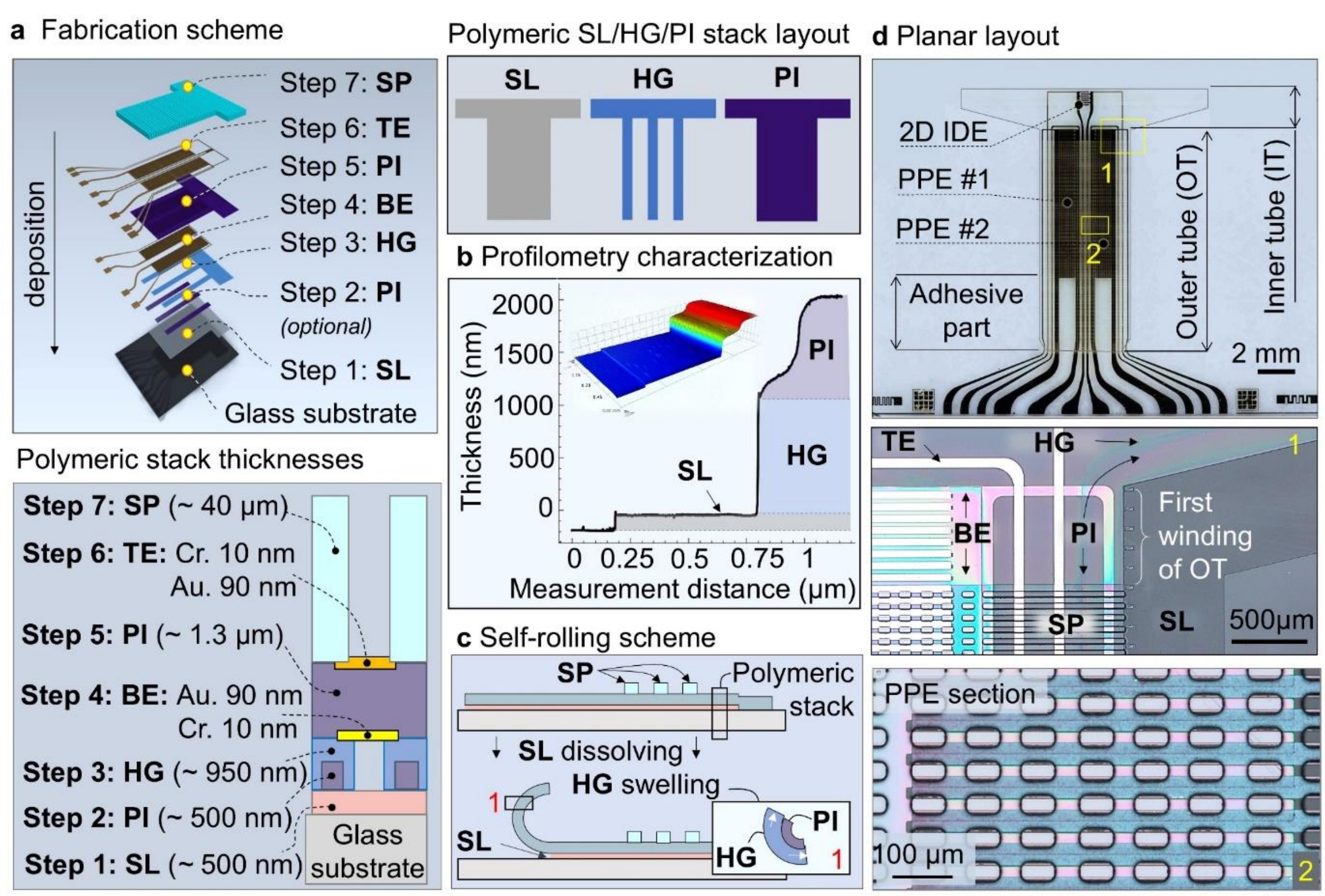

**Fig. S1** Fabrication process and structural layout of the coaxial Swiss-roll device. **a** Sequential thin-film deposition on a glass substrate, including sacrificial layer (SL), polyimide (PI), hydrogel (HG), bottom electrode (BE), top electrode (TE), and structural polymer (SP). Low panel: Cross-sectional schematic of the polymer stack with representative thicknesses: SL (~500 nm), PI (~500 nm, optional), HG (~950 nm), BE (Cr 10 nm / Au 90 nm), PI (~1.3 μm), TE (Cr 10 nm / Au 90 nm), and SP (~40 μm). Scale bars: 100 μm, 2 mm, and 500 μm. Right panel: schematic layout of polymeric SL/HG/PI stack layers. **b** Profilometry characterization with 3D map of SL/HG/PI polymeric stack. **c** Self-rolling mechanism: selective dissolution of the SL combined with HG swelling induces strain mismatch within the multilayer stack, resulting in autonomous tubular formation. **d** Planar device layout prior to rolling, showing the 2D interdigitated electrode (IDE), PPE sections, adhesive region, and the defined outer tube (OT) and inner tube (IT).

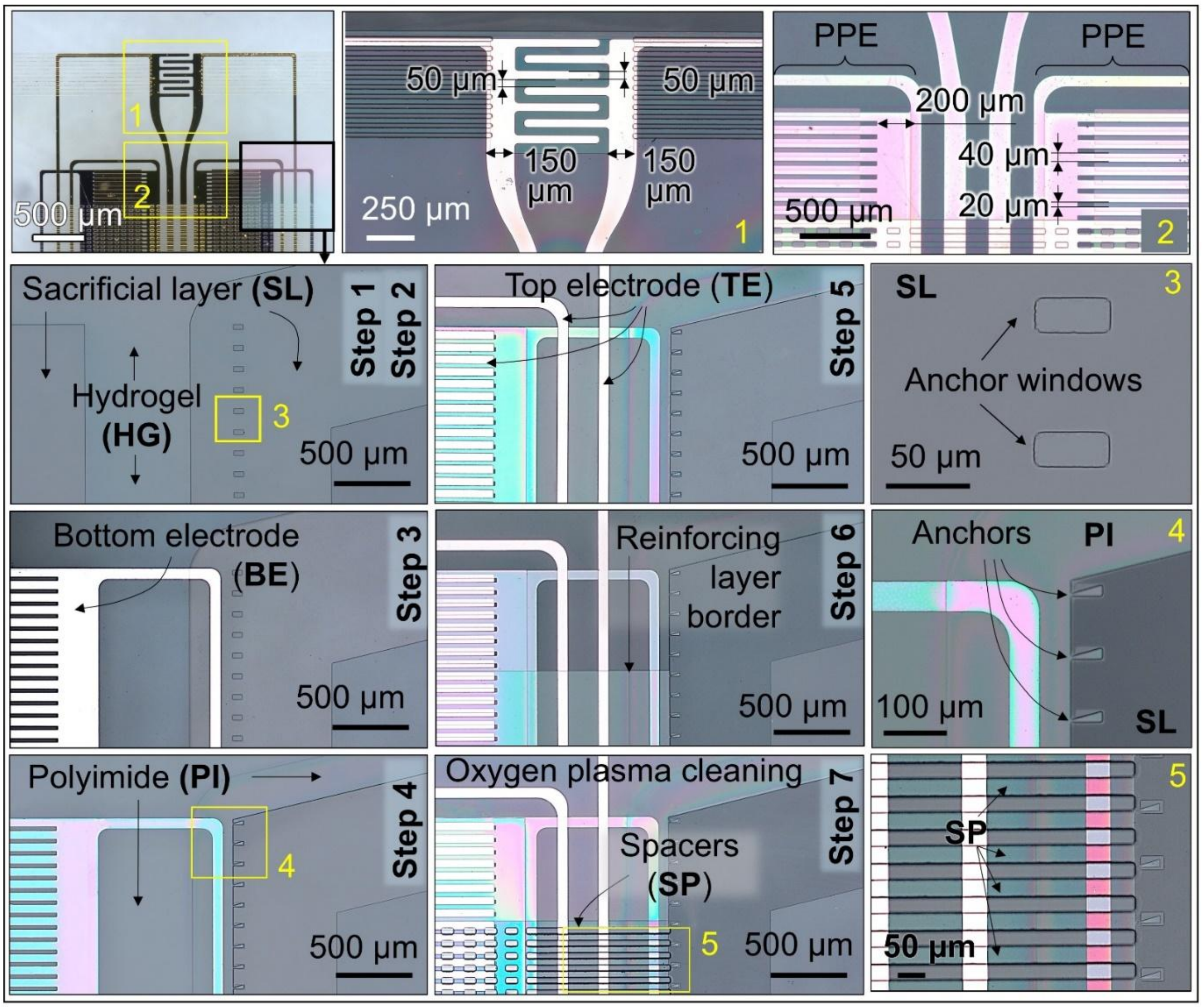

**Fig. S2** Detailed microfabrication sequence. Step-by-step photolithographic process for the polymeric nanomembrane stack: (Step 1) Sacrificial layer (SL) patterning. (Step 2) Hydrogel (HG) deposition and patterning. (Step 3) Sputtering and patterning of the bottom electrode (BE, Cr/Au). (Step 4) Spin-coating and patterning of the first polyimide (PI) layer. (Step 5) Sputtering and patterning of the top electrode (TE, Cr/Au). (Step 6) Deposition and patterning of a thicker PI reinforcing layer for mechanical stability. (Step 7) Oxygen plasma cleaning and final deposition/patterning of SU-8 spacer (SP) bars. Insets 1, 2: Detailed dimensions of the electrode structures. Insets 3, 4: Design and integration of lithographic anchors to prevent lateral rolling. Inset 5: Close-up view of the SU-8 spacer bars.

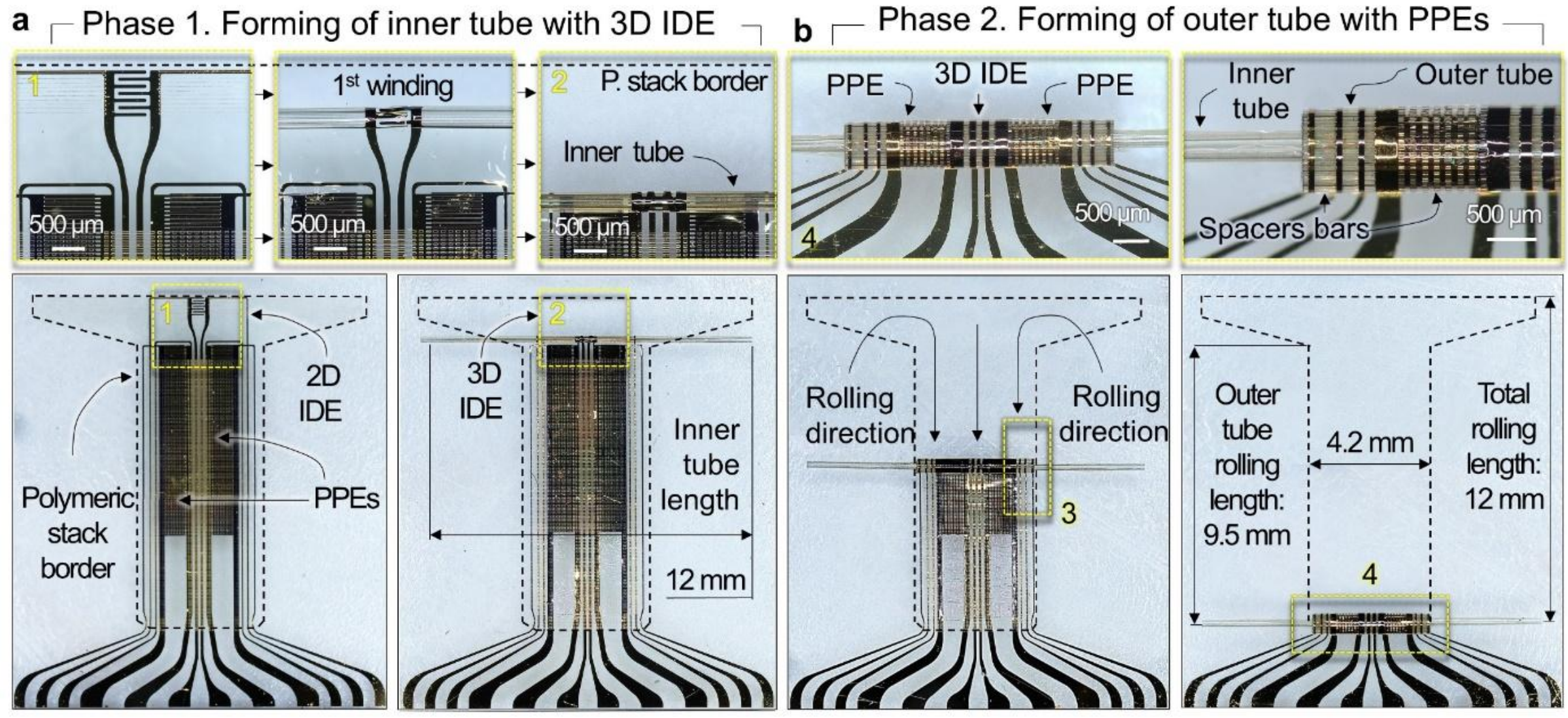

**Fig. S3** Sequential self-assembly of the 3D coaxial architecture. The strain-driven rolling process occurs in two distinct phases, guided by lithographic anchors and a polyimide (PI) reinforcing layer to ensure the correct sequence. **a** Phase 1: Inner tube formation. Overview image and corresponding zoomed views (Insets 1, 2) showing the initial rolling stage that transforms the 2D interdigitated electrodes (IDEs) into the 3D inner tube. **b** Phase 2: Outer tube formation. Overview image and corresponding zoomed views (Insets 3, 4) showing the subsequent rolling stage, where the outer structure encases the inner tube to form the complete coaxial architecture with integrated parallel-plate electrodes (PPEs). The final device has a total length of 12 mm.

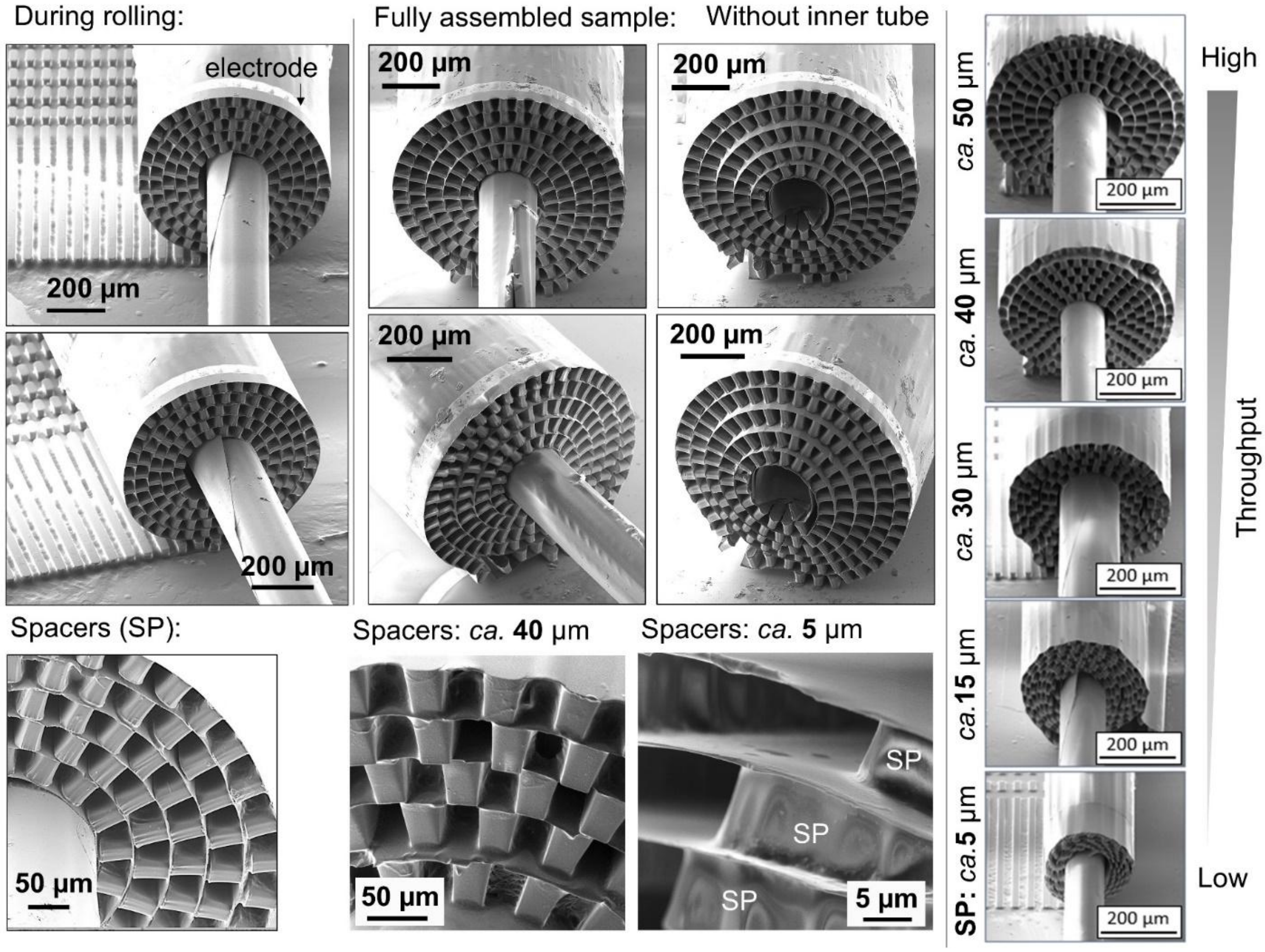

**Fig. S4** Scanning electron microscopy (SEM) images of self-assembled Swiss-roll coaxial platform with varying interwinding gaps. The top row shows the fabrication process, illustrating the rolling of a layered structure into a coaxial configuration (left), resulting in the fully assembled samples (centre). The fully assembled samples exhibit a consistent pattern of concentric rings around a central inner tube, with the interwinding gap (spacer) distance systematically varied (right). The bottom rows show higher magnification views of the spacers for samples with ca. 40 μm and ca. 5 μm gaps, respectively, highlighting the microstructural morphology. The right column presents a series of fully assembled samples with varying interwinding gaps: 5 μm (volume of outer tube is ~70 nL), 15 μm (volume of outer tube is ~200 nL), 30 μm (volume of outer tube is ~400 nL), 40 μm (volume of outer tube is ~550 nL), and 50 μm (volume of outer tube is ~700 nL), demonstrating a clear correlation between the spacer size and the resulting macroscopic structure. The gradient bar on the right indicates the relative throughput associated with each spacer size, with larger gaps yielding higher throughput.

**Fig. S5** Microfluidic integration and sealing of the 3D coaxial platform. **a** Sealing strategy. Optical micrographs showing the integration of the self-assembled coaxial structure within the PDMS device. The architecture is aligned in the main channel, and liquid PDMS is introduced through lateral channels to locally seal the structure, establishing isolated fluidic connections. **b** Programmable sequential filling. A series of optical micrographs demonstrating controlled, winding-by-winding fluid progression through the 3D architecture. The SU-8 spacer design creates discrete microfluidic channels within the windings, enabling precise control over the filled volume. This confirms the design provides addressable microfluidic access to the entire active surface area. Scale bars 50 µm.

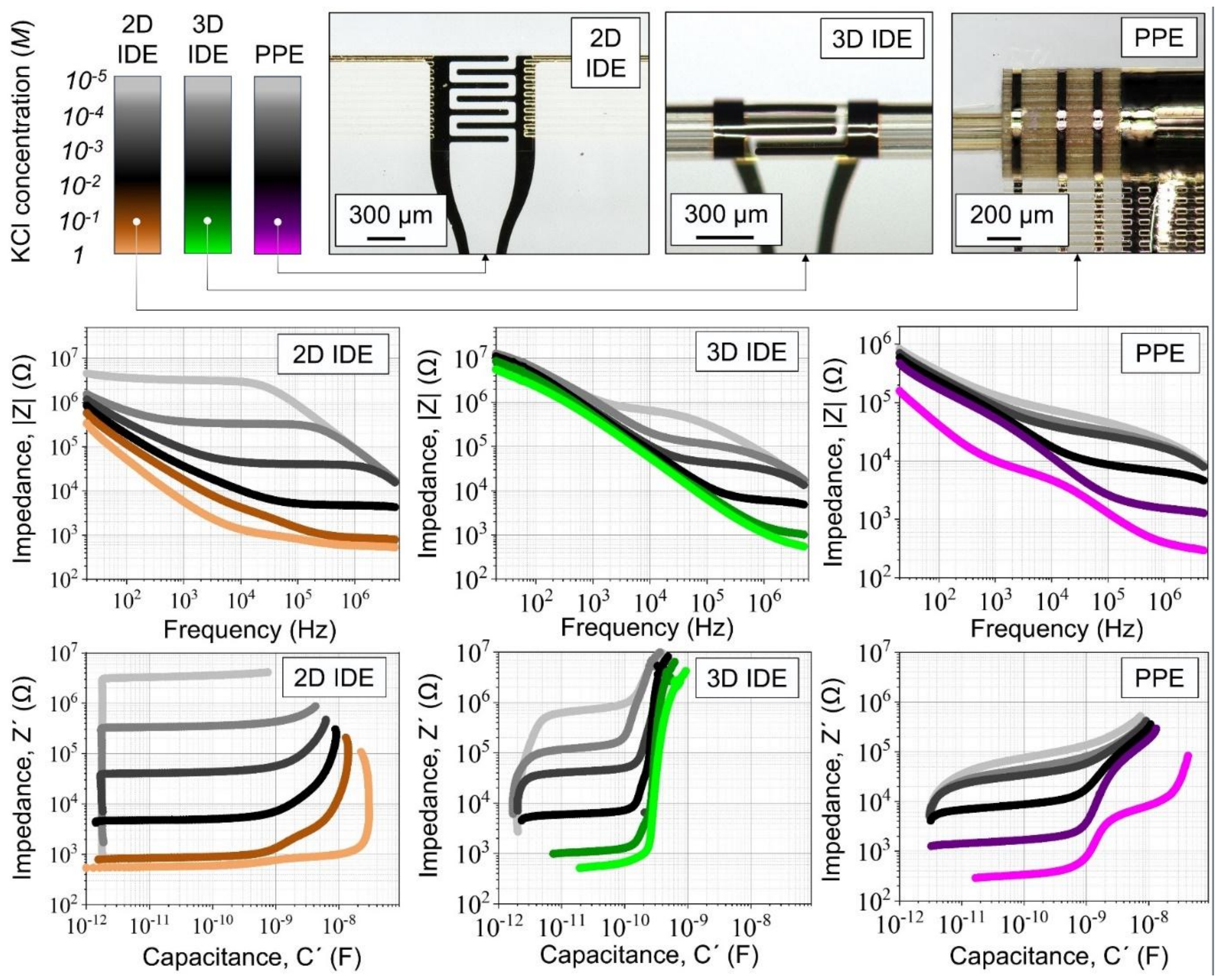

**Fig. S6** Geometry-dependent electrochemical performance of the integrated electrodes. Bode plots (top row, impedance magnitude |Z| vs. frequency) and capacitance spectra (bottom row) directly compare the three electrode geometries - planar 2D IDEs, 3D coaxial IDEs, and PPEs - across KCl solutions ($10^{-5}$ to 1 M). The data reveal a pronounced geometry effect: the 3D self-assembled structures (3D IDEs, PPEs) consistently outperform the 2D baseline, with PPEs providing the lowest impedance and most stable capacitance due to their large surface area, while 3D IDEs offer robust performance across all ionic strengths.

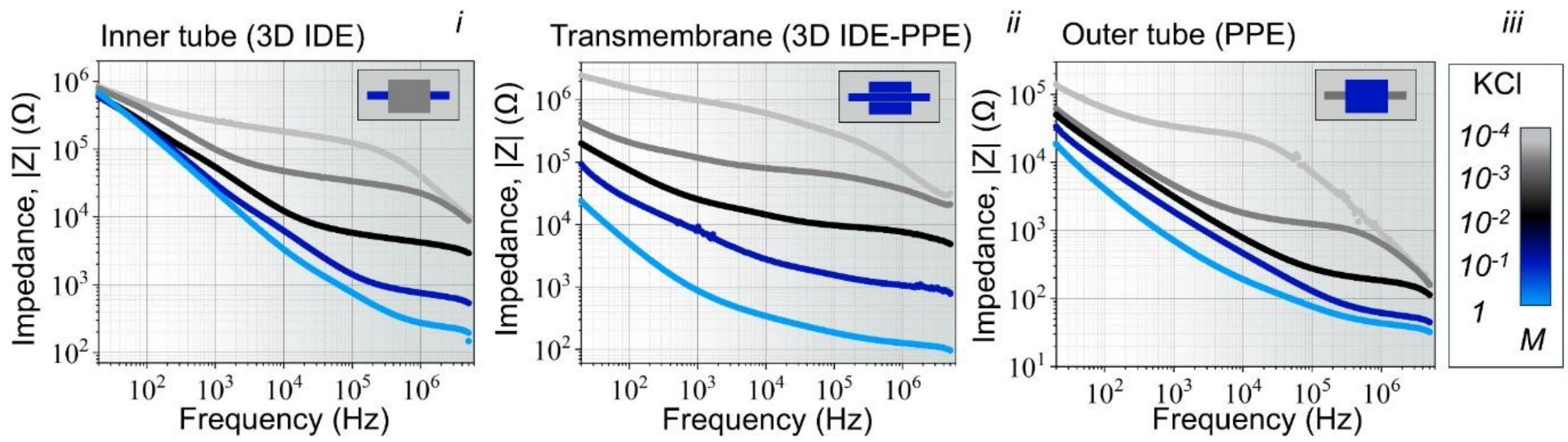

**Fig. S7** Bode plot analysis corresponding to Fig. 3c. Impedance magnitude (|Z|) versus frequency for the same measurements shown in the capacitance–impedance plots of Fig. 3c, featuring (*i*) 3D IDEs, (*ii*) 3D IDE-PPE pair, and (*iii*) PPEs across KCl solutions ($10^{-4}$ to 1 M). These Bode plots provide complementary frequency-domain analysis, quantitatively showing that PPEs maintain the lowest impedance magnitude across most frequencies, while 3D IDEs offer stable performance over the broadest frequency range.

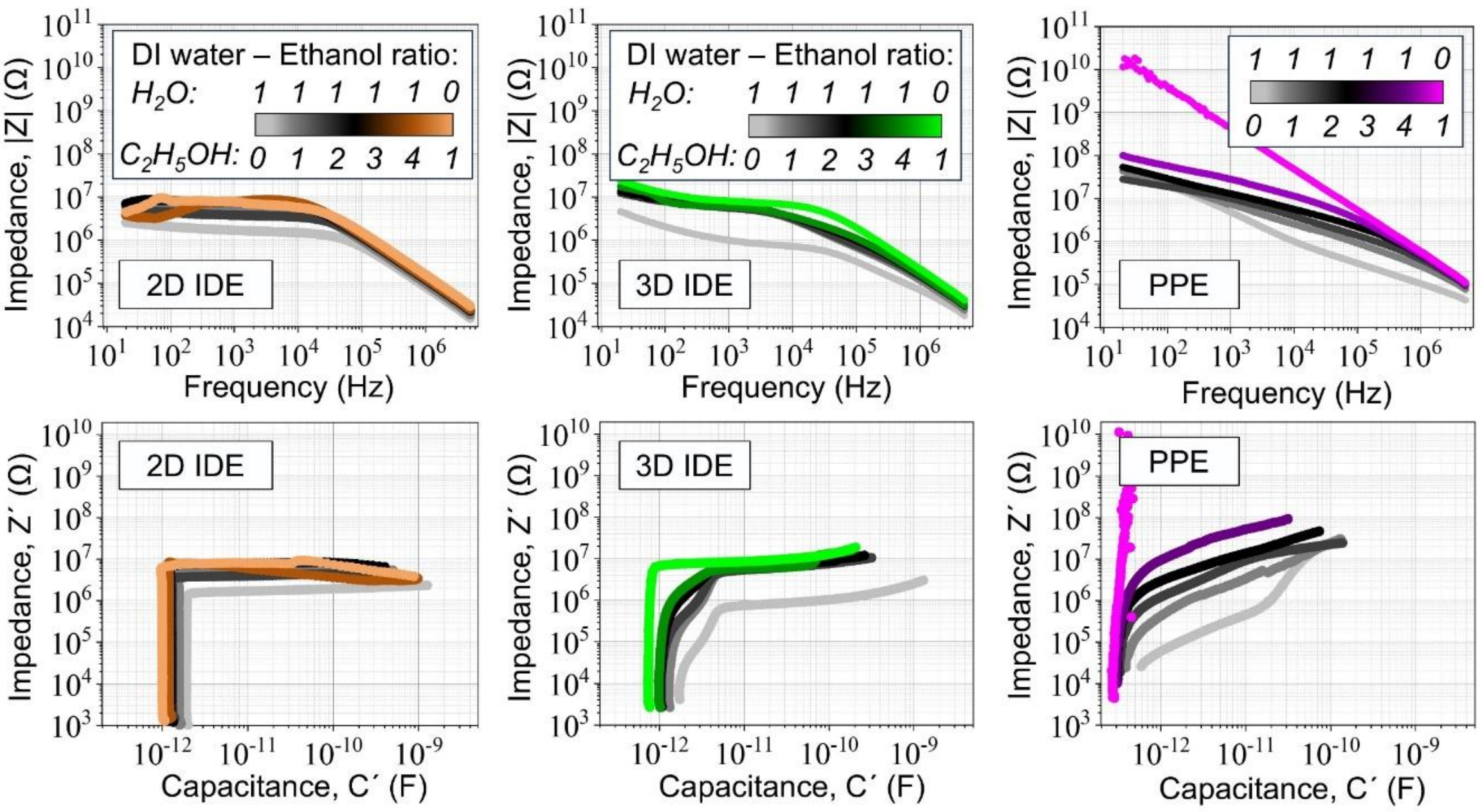

**Fig. S8** Dielectric sensitivity of electrode geometries. Bode plots (top row) and corresponding capacitance spectra (bottom row) for the 2D IDEs, 3D IDEs, and PPEs in ethanol–water mixtures of varying composition (pure water to pure ethanol). The data quantify the geometry-dependent dielectric response: PPEs act as highly sensitive dielectric sensors, with impedance varying by nearly three orders of magnitude, while the IDE geometries provide stable performance, making them robust for fluctuating environments. This demonstrates the capability to select electrode geometry based on the required application, whether sensing or stable operation.

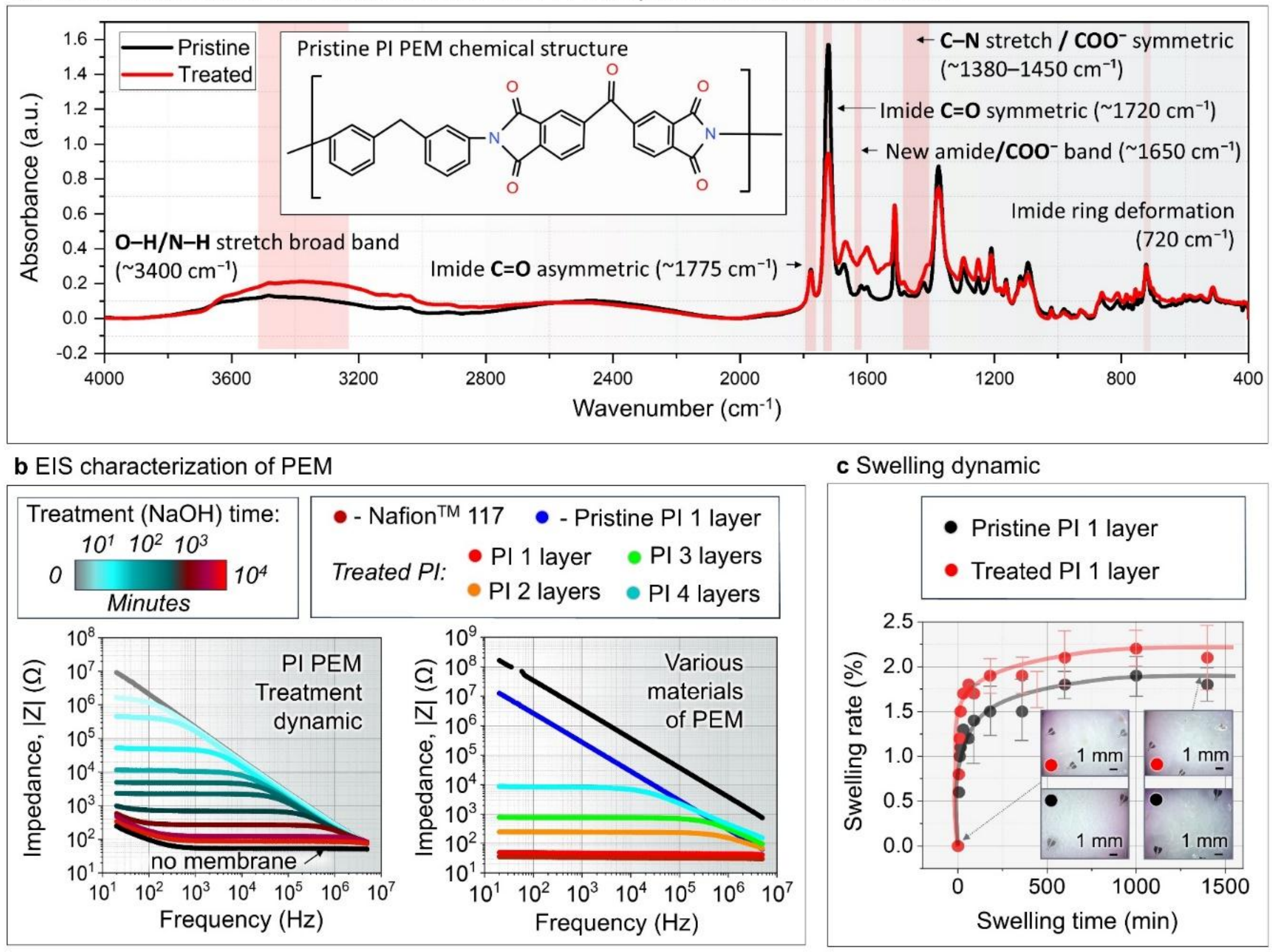

**Fig. S9** Chemical and electrochemical evidence for PI PEM tunability. **a** FTIR spectra of pristine (black) and NaOH-treated (red) PI membranes. The treatment reduces characteristic imide peaks (asymmetric C=O stretch at ~1775 $cm^{-1}$, symmetric at ~1720 $cm^{-1}$) and introduces new bands (broad O–H/N–H at 3200–3600 $cm^{-1}$, amide C=O at ~1650 $cm^{-1}$, $COO^{-}$ at ~1420 $cm^{-1}$), confirming the partial ring-opening into poly(amic acid) and carboxylate species that enhance hydrophilicity and proton conduction. **b** Electrochemical impedance spectroscopy (Bode plots, |Z| vs. frequency) provides the underlying data for the performance metrics in Fig. 4d. The plots show the systematic reduction of membrane resistance with NaOH treatment time (left, corresponding to Fig. 4d, ii) and benchmark the performance of the optimally treated PI against commercial standards (right, corresponding to Fig. 4d, iii). **c** Swelling dynamics of PI membranes. Time-resolved swelling of pristine (black) and NaOH-treated (red) PI membranes measured by optical tracking of metal markers over 24 h immersion in aqueous electrolyte. Pristine membranes exhibited 1.5–1.7% swelling, while treated membranes showed 2.1–2.3% swelling. The moderate expansion remains within the elastic limit of integrated gold electrodes, confirming mechanical compatibility.

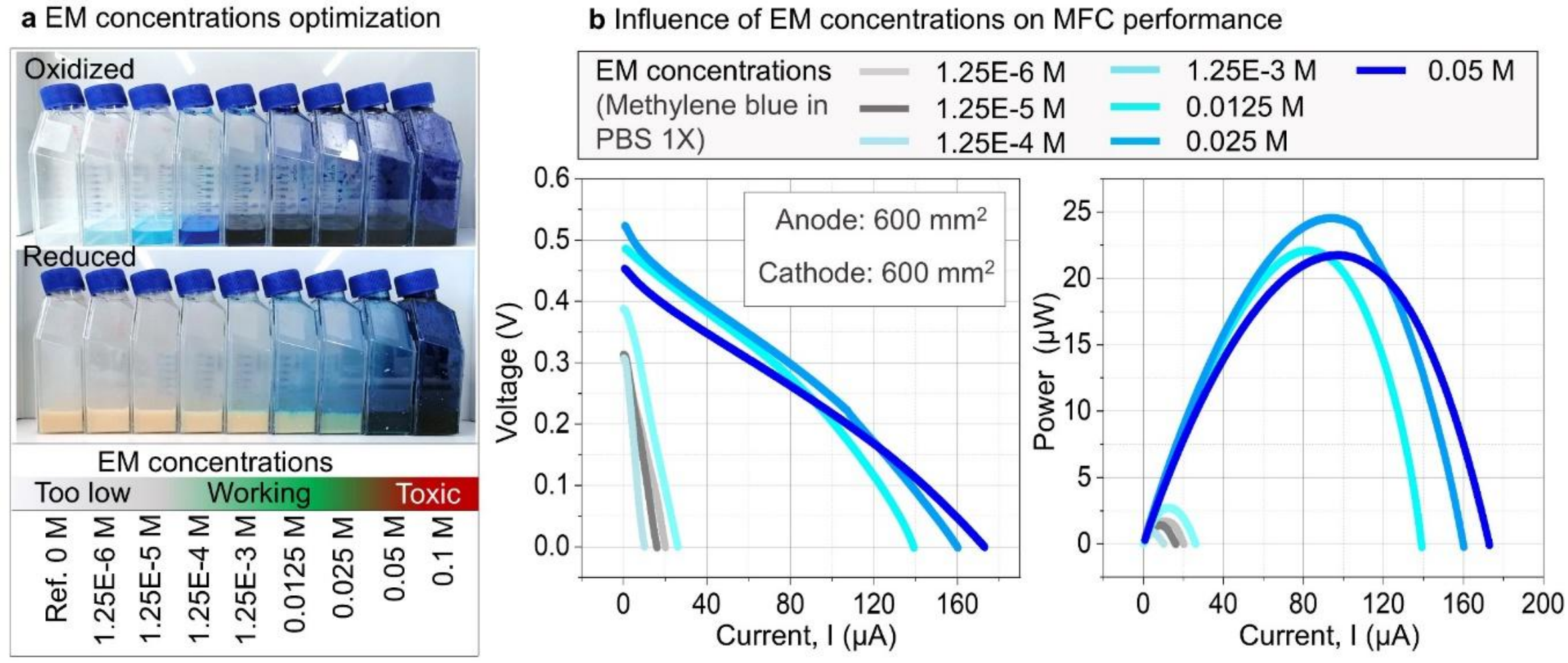

**Fig. S10** Influence of exogenous mediator concentration on system performance. **a** Photographs of methylene blue solutions, showing the oxidized (blue, top) and reduced (colorless, bottom) states across a concentration gradient. The visible color transition confirms active redox cycling within the operational range. **b** Polarization and power density curves quantifying the performance window. Optimal output occurs at mediator concentrations of $1.25 \times 10^{-3}$ M to 0.025 M. Performance declines at lower concentrations due to insufficient electron transfer and at higher concentrations ($\geq$0.05 M) due to cytotoxic effects.

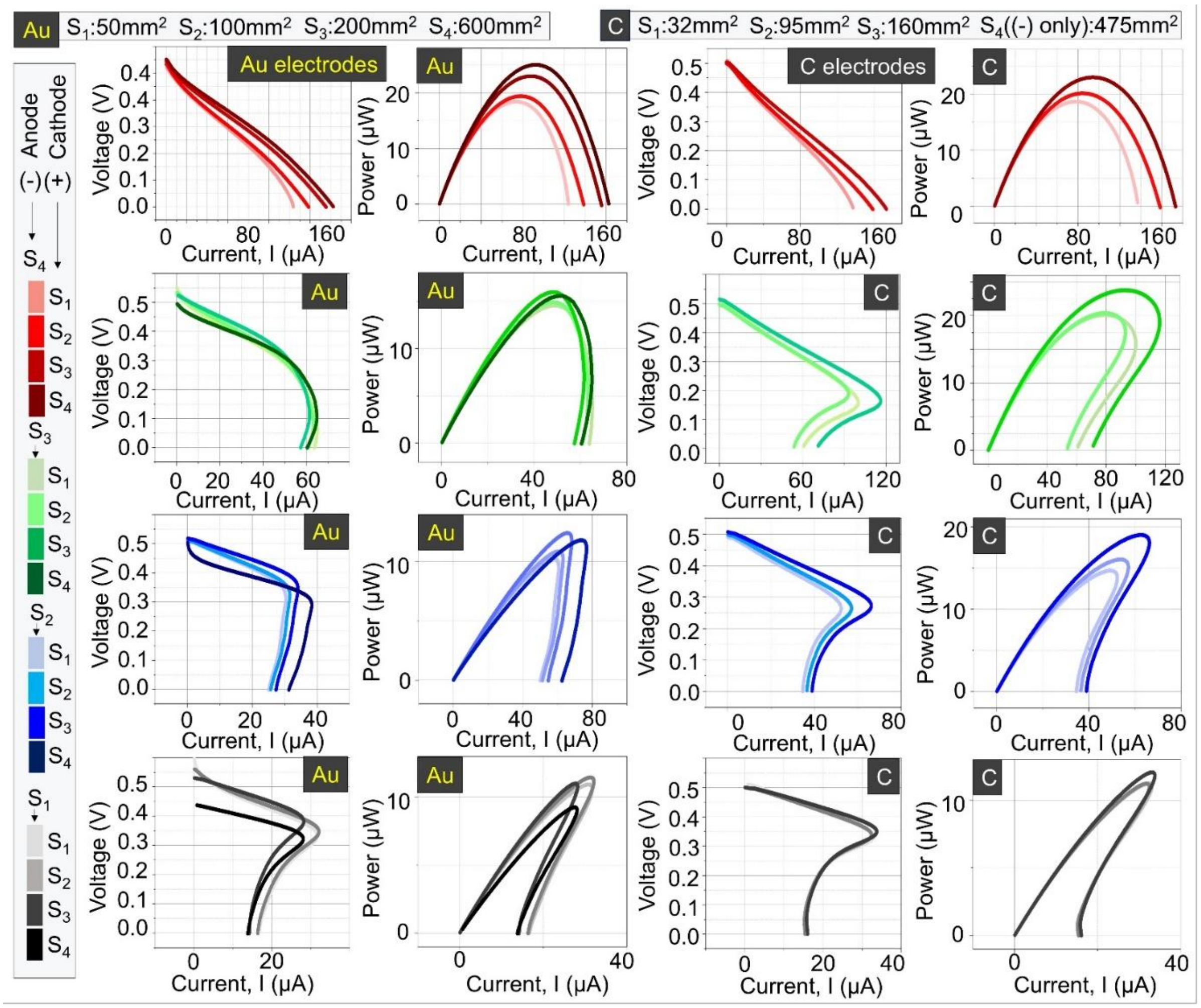

**Fig. S11** Electrode material and surface area optimization. Polarization and corresponding power density curves for gold (Au) and graphite (C) electrodes of increasing surface area (S1–S4). The data demonstrate that enlarging the anode surface area markedly improves current and power output, with Au electrodes consistently outperforming graphite. Cathode enlargement provided minimal performance gains. These results directly informed the selection of Au and the prioritization of anode area in the coaxial µMFC design.

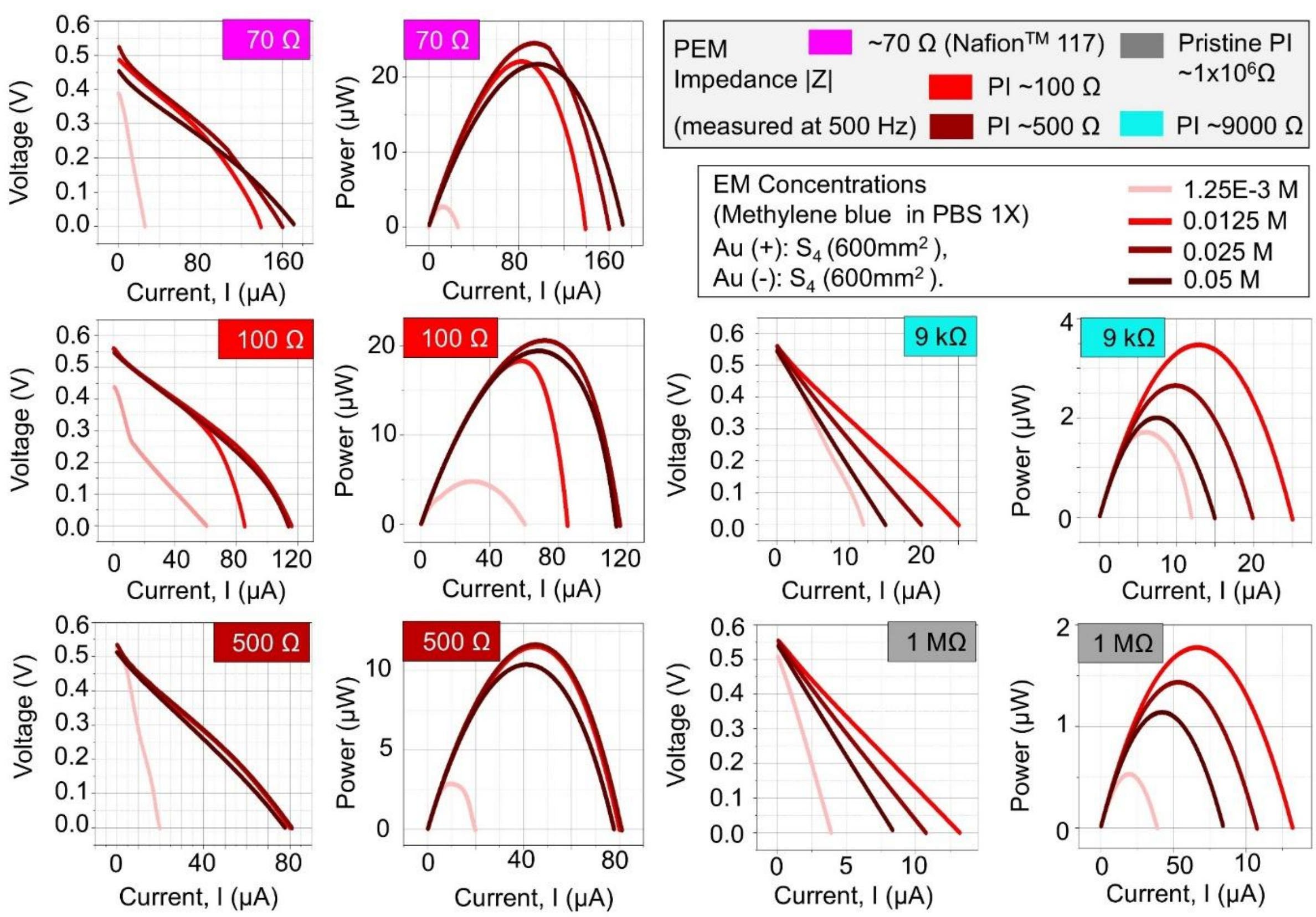

**Fig. S12** Independent optimization of PEM resistance and mediator concentration. Polarization and power density curves for PI proton-exchange membranes (PEMs) of varying resistance (70 Ω to 1 MΩ) across a range of methylene blue concentrations. The data demonstrate two independent design axes: (1) The optimal mediator concentration window (~0.0125–0.025 M) is consistent across all PEM resistances. (2) Within this optimal window, the absolute performance is exclusively dictated by PEM resistance, with low-resistance membranes (~70-250 Ω) enabling the highest power output.

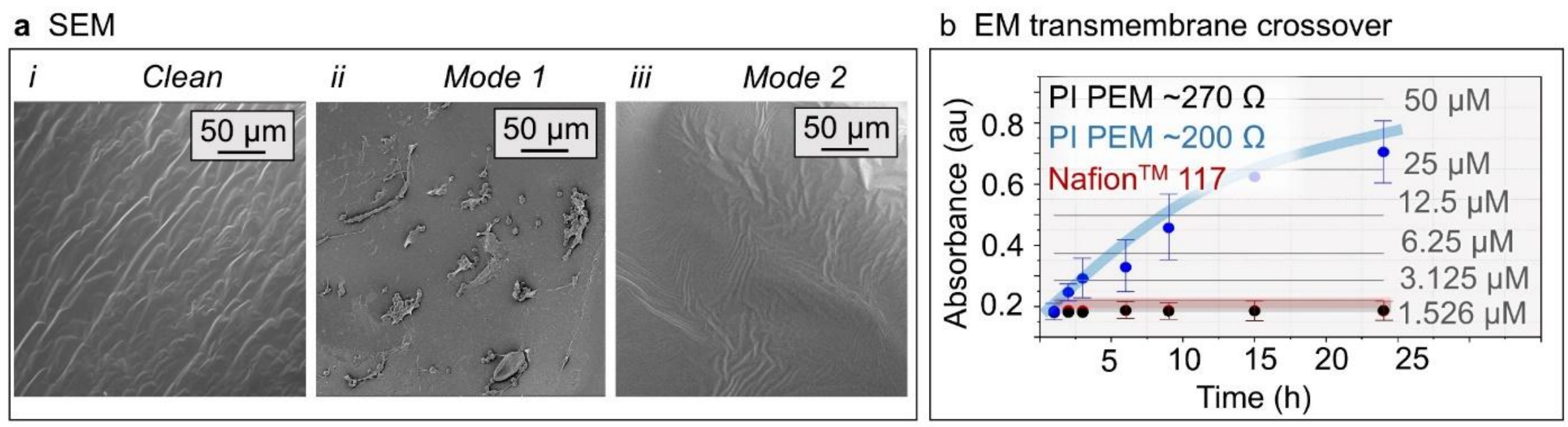

**Fig. S13.** Biofouling and mediator crossover characterization. **a** SEM images of PEM surfaces under clean conditions (*i*) and after exposure to Mode 1 (*ii*) and Mode 2 (*iii*). Mode 1 shows extensive surface coverage and biomass aggregates, whereas Mode 2 retains the underlying membrane topography, indicating absence of biofilm formation. **b** EM crossover over 24 h for PI PEM (~200 Ω), PI PEM (~270 Ω), and Nafion™ 117; optimized PI PEM (~270 Ω) blocks mediator transport below detection limit.

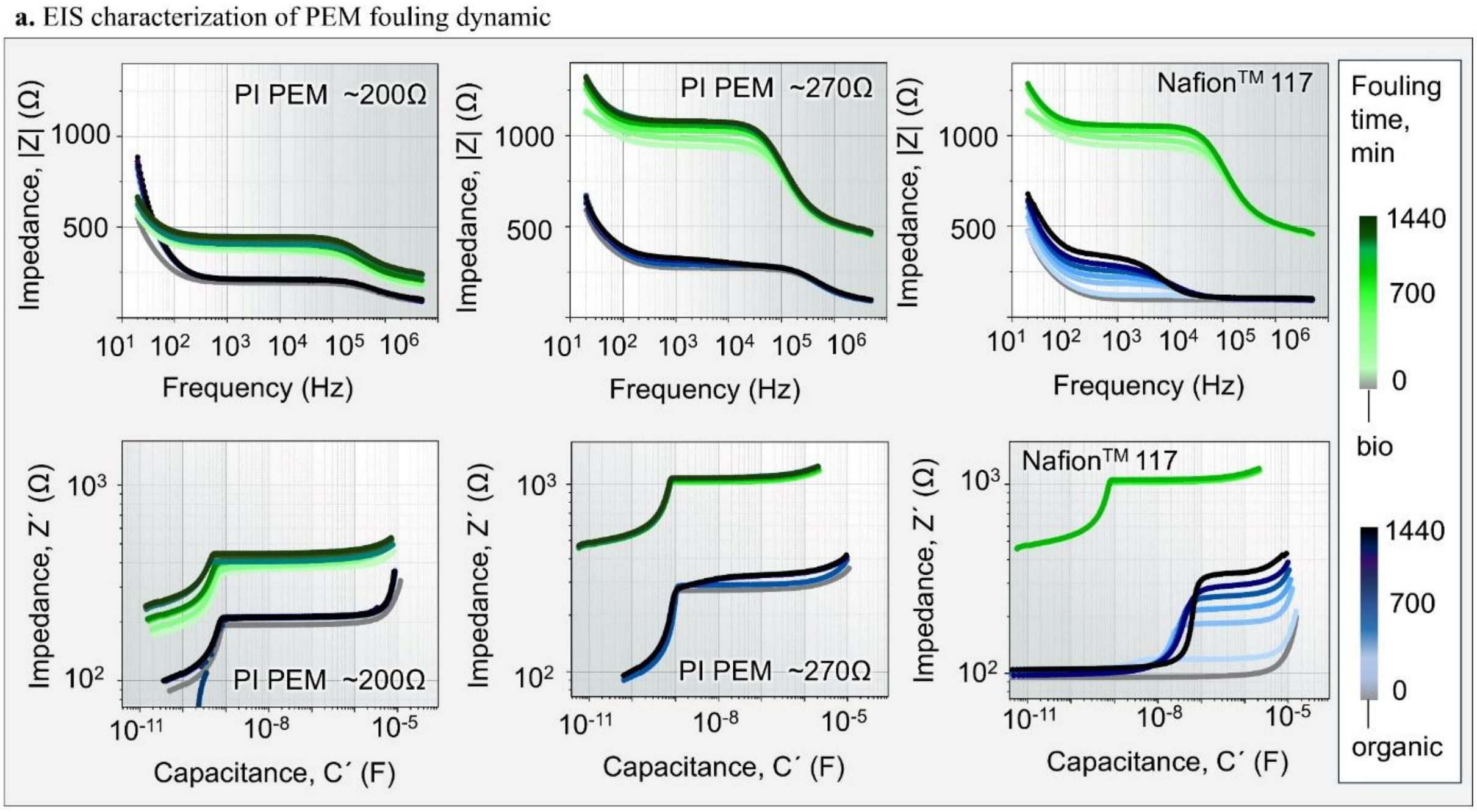

**Fig. S14** EIS characterization of PEMs. Bode plots (top row, impedance magnitude |Z| vs. frequency) and capacitance spectra (bottom row) directly compare the three PEMs: PI PEM ~200 Ω (left), PI PEM ~270 Ω (central) and Nafion™ 117 PEM.

**Supplementary Video V1–5**

**Video V1: Independent and stable fluidic operation of inner and outer channels.**

The video demonstrates the integrity of the integrated microfluidic platform, showing independent injection of orange (inner channel) and green (outer channel) fluid streams. Flow is pulsed in forward and reverse directions, confirming complete isolation between the two coaxial compartments with no observable mixing, as illustrated in the connection scheme in Fig. 3b. (Playback speed: 2x real-time)

**Video V2: Controlled fluidic transfer between inner and outer channels.**

This video illustrates the designed fluidic routing capability. A red dye introduced into the inner tube transfers to the outer tube via an external interconnection, subsequently flowing in the opposite direction within the outer coaxial channel. This demonstrates programmable fluidic control within the sealed microsystem. (Playback speed: 2x real-time)

**Video V3: Independent filling of inner and outer channels in an open configuration.**

Deionized water is shown independently filling the inner and outer tubes of the Swiss-roll structure, confirming the successful self-assembly of two distinct, continuous microfluidic channels. (Playback speed: 2x real-time)

**Video V4: Sequential and controlled filling of the interwinding channels.**

In an open configuration, deionized water fills the interstitial channels between the windings in a sequential manner, as guided by the SU-8 spacer design (related to Supplementary Fig. S5b). This demonstrates controlled fluidic access to the entire 3D geometry and the potential for localized fluidic addressing. (Playback speed: 2x real-time)

**Video V5: Visualization of sequential interwinding filling with colored fluid.**

A colored dye visually enhances the demonstration of the sequential filling process, clearly showing the controlled progression of fluid through the interwinding channels one by one, confirming efficient fluidic access to the coiled architecture. (Playback speed: 2x real-time)